\documentclass[twocolumn]{aastex63}

\usepackage{hyperref}
\usepackage{amsmath}
\usepackage{amssymb}
\usepackage{graphicx}

\newcommand{\teff}{$T_{\rm eff}$}
\newcommand{\logg}{$\log g$}
\newcommand{\vt}{$\xi_{t}$}

\newcommand{\AB}[2]{$\mbox{[#1/#2]}$}
\newcommand{\feh}{\AB{Fe}{H}}

\begin{document}

\title{Probing Abundance Variations among Multiple Stellar Populations\\ in the Metal-Poor Globular Cluster NGC 2298 using Gemini-South/GHOST

\footnote{Based on observations obtained at the international Gemini Observatory, a program of NSF NOIRLab, which is managed by the Association of Universities for Research in Astronomy (AURA) under a cooperative agreement with the U.S. National Science Foundation on behalf of the Gemini Observatory partnership: the U.S. National Science Foundation (United States), National Research Council (Canada), Agencia Nacional de Investigación y Desarrollo (Chile), Ministerio de Ciencia, Tecnología e Innovación (Argentina), Ministério da Ciência, Tecnologia, Inovações e Comunicações (Brazil), and Korea Astronomy and Space Science Institute (Republic of Korea).}}

\correspondingauthor{Avrajit Bandyopadhyay}
\email{abandyopadhyay@ufl.edu}

\author[0000-0002-8304-5444]{Avrajit Bandyopadhyay}
\affiliation{Department of Astronomy, University of Florida, Bryant Space Science Center, Gainesville, FL 32611, USA}
\affiliation{Joint Institute for Nuclear Astrophysics -- Center for Evolution of the Elements, USA}

\author[0000-0002-8504-8470]{Rana Ezzeddine}
\affiliation{Department of Astronomy, University of Florida, Bryant Space Science Center, Gainesville, FL 32611, USA}
\affiliation{Joint Institute for Nuclear Astrophysics -- Center for Evolution of the Elements, USA}

\author[0000-0003-4479-1265]{Vinicius M. Placco}
\affiliation{NSF NOIRLab, Tucson, AZ 85719, USA}
\affiliation{Joint Institute for Nuclear Astrophysics -- Center for Evolution of the Elements, USA}

\author[0000-0002-2139-7145]{Anna Frebel}
\affiliation{Department of Physics \& Kavli Institute for Astrophysics and Space Research, Massachusetts Institute of Technology, Cambridge, MA 02139, USA}
\affiliation{Joint Institute for Nuclear Astrophysics -- Center for Evolution of the Elements, USA}

\author[0000-0001-5200-3973]{David S Aguado}
\affiliation{Instituto de Astrof\'{i}sica de Canarias, V\'{i}a L\'{a}ctea, 38205 La Laguna, Tenerife, Spain}
\affiliation{Universidad de La Laguna, Departamento de Astrofísica, E-38206 La Laguna, Tenerife, Spain}

\author[0000-0001-5107-8930]{Ian U. Roederer}
\affiliation{Department of Physics, North Carolina State University; Raleigh, NC 27695, USA}
\affiliation{Joint Institute for Nuclear Astrophysics -- Center for Evolution of the Elements, USA}

\begin{abstract}

Studying the abundances in metal-poor globular clusters is crucial for understanding the formation of the Galaxy and the nucleosynthesis processes in the early Universe. We observed 13 red giant stars from the metal-poor globular cluster NGC 2298 using the newly commissioned GHOST spectrograph at Gemini South. We derived stellar parameters and abundances for 36 species across 32 elements, including 16 neutron-capture elements. We find that the stars exhibit chemical anomalies among the light elements, allowing us to classify them into first-generation (8 stars) and second-generation (5 stars). We derive a mean cluster metallicity of [Fe/H]=$-$1.98 $\pm$ 0.10 with no significant variation among cluster members. Most $\alpha$- and Fe-peak elements display low star-to-star abundance dispersion, with notable exceptions for Sc, Ni, and Zn for which the dispersions in Sc vary significantly between stars from different generations to 2$\sigma$ levels. Similarly, among the neutron-capture elements, we observed considerable differences in dispersion for Sr and Eu among the first and second generation stars to 2$\sigma$ levels. We also confirm an intrinsic scatter beyond observational uncertainties for several elements using a maximum likelihood approach among stars from different generations. Additionally, we note an increase in [Sr/Eu] and [Ba/Eu] with [Mg/Fe] in first-generation stars indicating correlations between the productions of light $r$-process and Mg. We find the universal $r$-process pattern, but with larger dispersions in the main $r$-process than the limited-$r$ elements. These differences in abundance dispersion, among first- and second-generation stars in NGC 2298, suggest complex and inhomogeneous early chemical enrichment processes, driven by contributions from multiple nucleosynthetic events, including massive stars and rare $r$-process events. 

\end{abstract}

\keywords{nucleosynthesis ---  stars: abundances ---  stars: Population II --- abundances: globular clusters --- stars: atmospheres --- stars: fundamental parameters}

\section{Introduction} \label{sec:intro}

Globular clusters (GCs) are among the oldest known stellar populations in the Milky Way, serving as crucial repositories and providing vital insights into the epoch of structure formation in the early universe \citep{kimm2016,elme2018,forbes2018,is2023}. The study of GCs is essential for understanding the initial conditions of galaxy formation, as these clusters retain chemical signatures from the early interstellar medium \citep{james2004,bailin2009,Leaman_2012,bekki17,lucia2024}. Traditionally viewed as simple stellar systems consisting of a single stellar population \citep{krafta,kraftb}, advancements in photometric precision and spectroscopic analysis have revolutionized our understanding, revealing that many well-studied GCs harbor multiple stellar populations \citep{carretta2009,gratton2012,meszaros2020}. These populations are characterised by significant variations in light element abundances, such as C, N, O, Na, Mg, and Al, often exhibiting variations up to 1.0 dex \citep{sillis2010,dotter2014,Roe2015,cummings2017,simpson2017,bangc}. Understanding these variations is essential for understanding the complexities of stellar evolution and, hence, the formation of GCs themselves \citep{gratton2004,beasley2020}.

The light element anomalies such as the the Na-O and Mg-Al anticorrelations observed in globular clusters \citep{gratton2001}, suggest that different stellar populations within GCs have undergone varying degrees of internal enrichment \citep{carretta2010x,gratton2019,dantona2019}. For instance, the Na-O anticorrelation indicates that some stars have experienced proton-capture nucleosynthesis at high temperatures, leading to the depletion of O and the enhancement of Na \citep{Lee2010,gratton2011,munoz2017,car2019}. Similarly, the Mg-Al anticorrelation points to the operation of the Mg-Al cycle, where Mg is depleted and Al is enhanced in certain stellar environments \citep{den97,pan2017,mucci2018,baeza2022}. These chemical anomalies, uniformly present across all stellar evolutionary stages in several GCs, ruled out mixing within the star or contamination from secondary sources \citep{gratton2004}. The theory of multiple populations was confirmed by \citet{brag2010}, who linked these inhomogeneities (e.g., Mg-Al and Na-O anticorrelations) to distinct photometric sequences \citep{piotto2007,milone2012}. This finding established these inhomogeneities as signatures of multiple stellar generations, resulting from the recycling of advanced hydrogen-burning products from the first generation of stars in GCs, into subsequent stellar generations \citep{gratton2012,bastian2017,milone2022}. 
The presence of second-generation stars with distinct chemical signatures indicates that GCs underwent self-enrichment, likely through the ejecta of massive stars and asymptotic giant branch stars \citep{kiml18,dantona2019,don2022}. 

Existing theoretical models continue to struggle with simultaneously accounting for the observed light element dispersion, the apparent homogeneity of heavier elements, and other spatial and dynamical characteristics in GCs \citep{don2022}. However, recent findings by \citet{kirby23} in M92 have highlighted a significant dispersion in heavy \emph{r}-process elements among first-generation stars, suggesting that these stars may have formed from gas already enriched by an \emph{r}-process event, prior to  and faster than the expected mixing timescale within the cluster. This observation challenges the prevailing results that neutron-capture elements should not exhibit star-to-star dispersion in "normal" GCs. Understanding the spread and abundance levels of \emph{r}-process elements in stars from different generations within metal-poor GCs can provide valuable constraints on the mixing and formation timescales of the different populations \citep{Bekki19,Zevin19} . Furthermore, establishing any correlations between light elements and \emph{r}-process elements in this cluster would be helpful for our understanding of star-to-star neutron-capture element variations, which are currently believed to be confined to the most massive and peculiar GCs \citep{Bekki2007,Roederer2011, Johnson2015,kirby23}. These insights are not only critical for reconstructing the early stages of the Milky Way but also for informing models of galaxy formation and evolution on a cosmological scale \citep{bastian2017}.

NGC 2298 is a metal-poor, low-mass halo globular cluster with [Fe/H] = $-$1.98 that presents an intriguing target for detailed spectroscopic studies. As one of the smallest globular clusters (GCs) in the Milky Way \citep{pas2009}, NGC 2298 is currently undergoing accretion into the Galaxy \citep{martin2004,bal2011,julio2018}  and has been kinematically linked to the merged dwarf galaxy Gaia-Enceladus \citep{massari19}. Its low metallicity and position in the outer halo with an apocentric distance of 16.44 Kpc \citep{baeza2022} suggest that its members are less likely to have been contaminated by disruptive events, thereby preserving early supernova signatures \citep{janes88}. Previous spectroscopic observations have provided preliminary data \citep{mc92,baeza2022} but a comprehensive and systematic abundance study is still lacking. \citet{baeza2022} indicated the presence of multiple populations in NGC 2298 based on the Mg-Al anticorrelation, highlighting the importance of deriving abundances for a large number of species in the member stars to fully characterize the GC and investigate the other associated trends. 

The primary goal of this study is to conduct a detailed and homogeneous spectroscopic analysis of 13 giant stars in NGC 2298 based on GHOST high-resolution spectra taken with the Gemini South telescope \citep{ghost}. By deriving abundances of 32 elements including 16 neutron-capture elements of the brightest cluster members, we aim to test the hypothesis of abundance variations among heavy elements, particularly the $r$-process, and their relationship to light elements across multiple populations.
In this paper, we present the target selection and data collection in Section 2. The analysis, including stellar parameters and derivation of the chemical abundances, is described in Section 3. Our results and associated discussion are presented in Section 4.

\section{Target selection and observational data} \label{sec:data}

\begin{figure*}
\centering
\includegraphics[scale=0.45]{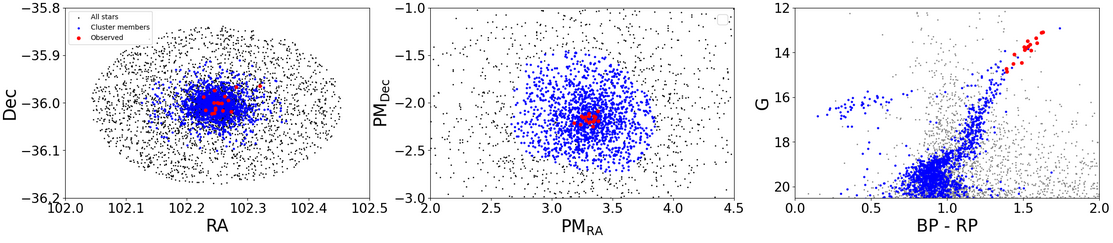}
\caption{Left: The distribution of the NGC 2298 marked in black along RA and DEC. The blue filled circle indicates the stars with cluster membership probability $\geq$ 0.99 with stringent cuts on parallax and proper motions along RA and Dec. Middle: The proper motions along RA and Dec for the sample of stars. The red filled circles mark the finally chosen spectroscopic targets. Right: The CMD for NGC 2298 with the same colors as in the left panel. The selected samples marked in red are the brightest bonafide cluster members.}
\label{targetselection}
\end{figure*}

For our target selection within NGC 2298, we utilized the Galactic globular cluster catalog by \citet{vasiliev2021}. The tidal radius, obtained from \citet{Moreno2014}, was projected onto the sky plane to identify cluster candidates. We selected candidates with membership probabilities exceeding 99\%, as shown in Figure~\ref{targetselection}. Gaia detections within a 10\arcmin\ radius of the cluster center were considered, excluding stars with Renormalised Unit Weight Error (RUWE) $>$ 1.2 and astrometric excess noise $>$ 1.3 mas. To eliminate field stars, we excluded those falling outside the 3$\sigma$ limits of the mean parallax of NGC 2298. We considered only the stars with proper motion errors within 3$\sigma$ of the mean proper motion for the cluster, depicted in blue in Figure~\ref{targetselection}.
Finally, to ensure the feasibility of the planned GHOST observations, we selected spectroscopic RGB stars brighter than 'Gaia' G $<$ 16.0 from the stars closely matching the RGB of the cluster's Color-magnitude diagram. Combined, these selection criteria yielded 13 stars suitable for the spectroscopic follow-up with GHOST.

Observations for 13 stars were carried out with the Gemini South Telescope in December 2023, employing the GHOST spectrograph\footnote{The data was observed in the Fast Turnaround programme for the dedicated call for shared-risk observations using GHOST (GS-2023B-FT-210, PI: AB)} \citep{Kalari2024}. Nine targets could be observed in dual mode, following the criteria for dual mode observations with GHOST, to minimise the observing time, while four targets had to be observed in the standard single fiber mode. The targets were observed using the standard high-resolution mode, providing a spectral resolution of 56,000 and covering the wavelength range from 3500 to 10,000 \AA. Each target was observed for an exposure time of 1800 seconds, yielding a signal-to-noise ratio (SNR) of approximately 45 at 4500 \AA.

The data were reduced with the DRAGONS data reduction software\footnote{https://dragons.readthedocs.io/} customized for GHOST data \citep{dragons}. The DRAGONS Recipe System efficiently handles both interactive and automated data reductions by dynamically associating inputs with the most suitable recipes \citep{Placco2024RNAAS}. The reduction steps included bias subtraction, flat-field corrections, wavelength calibration, sky subtraction, barycentric correction, and extraction of individual orders. The final 1D spectra are generated using a variance-weighted stitching of the spectral orders.

\begin{table*}
    \centering
    \caption{Observational Details of the Target Stars}
    \label{tab:ident}
    
    \begin{tabular}{lccccccccccccrr} 
        \hline
        Name & R.A. &Dec & $g$ mag &  Exp. time &Obs mode & SNR (5000 \AA) & \multicolumn{1}{r}{RV$_\mathrm{Gaia}$} &\multicolumn{1}{r}{eRV$_\mathrm{Gaia}$}  & \multicolumn{1}{r}{RV$_\mathrm{helio}$} &\multicolumn{1}{r}{eRV$_\mathrm{helio}$} \\ 
		\hline
NGC2298-11 &06:49:04.24 &$-$35:59:39.14 &14.4 &1800 &Single &36 &155.5 &4.69 &145.7 &2.4\\
NGC2298-12 &06:48:58.72 &$-$36:00:00.27 &13.7 &1800 &Single &39 &$-$ &147.6 &1.9\\  
NGC2298-13 &06:49:00.66 &$-$36:00:03.37 &13.4 &1800 &Single &42 &155.7 &3.7 &153.5 &2.8\\
NGC2298-14 &06:48:59.22 &$-$36:00:42.11 &13.7 &1800 &Single &38 &158.1 &6.6 &150.5 &3.4\\
NGC2298-581 &06:49:16.77 &$-$35:5750.48 &14.1 &1800 &Dual &36 &146.9 &3.3 &148.1 &2.9\\ 
NGC2298-582 &06:49:02.12 &$-$36:00:59.82 &13.6 &1800 &Dual &39 &143.6 &3.3 &146.0 &2.6\\
NGC2298-591 &06:48:58.86 &$-$35:58:26.30 &13.9 &1800 &Dual &36 &147.8 &3.9 &147.2 &2.8\\ 
NGC2298-592 &06:48:58.89 &$-$36:01:16.51 &13.1 &1800 &Dual &43 &150.3 &1.4 &150.5 &2.2\\
NGC2298-601 &06:49:05.57 &$-$36:01:05.05 &14.5 &1800 &Dual &34 &150.0 &4.9 &142.6 &3.6\\
NGC2298-602 &06:48:54.60 &$-$35:59:15.17 &13.9 &1800 &Dual &38 &142.8 &2.9 &147.8 &2.2\\
NGC2298-981 &06:49:07.35 &$-$35:58:01.34 &13.6 &1800 &Dual &41 &145.3 &2.4 &145.8 &2.8\\
NGC2298-611 &06:49:03.01 &$-$35:59:09.96 &13.8 &1800 &Dual &38 &146.2 &4.0 &146.6 &3.3\\
NGC2298-612 &06:48:55.36 &$-$36:00:56.62 &13.5 &1800 &Dual &41 &151.2 &2.7 &150.9 &2.8\\
	\hline
	\end{tabular}
\end{table*}

\section{Analysis} \label{sec:floats}

\subsection{Radial velocities}

Heliocentric radial velocities (RV$_{\mathrm{helio}}$) were calculated using the \texttt{rvcorrect} package within \texttt{PYRAF}. The results  are provided in Table~\ref{tab:ident}. The RVs for all 13 stars have a a mean RV of 148.2 km s$^{-1}$ which is consistent with the GC's system velocity of 148.9 km s$^{-1}$. For the majority of stars with corresponding $Gaia$ RVs, our results exhibit strong consistency, with a mean deviation of 2.2 km s$^{-1}$ and a standard deviation of 2.75 km s$^{-1}$. None of the stars show indications of being in a binary system. 

\subsection{Stellar parameters}

Stellar atmospheric parameters for our sample stars (effective temperature \teff; surface gravity \logg, metallicity \feh, and microturbulent velocity, \vt) were derived from measurements of equivalent widths (EW) of Fe I and Fe II lines. The equivalent widths of the Fe lines were measured by fitting Gaussian line profiles to the spectral absorption features using the software \texttt{SMHr}. Initial LTE parameters were estimated using the LTE radiative transfer code \texttt{MOOG} \citep{sneden1973}, including Rayleigh scattering treatment \citep{sobeck2011}\footnote{https://github.com/alexji/moog17scat}. We employed 1D, LTE stellar atmospheric ATLAS models from \citet{castelli2004}, with a standard $\alpha$-element enhancement of [$\alpha$/Fe] = +0.4.

Initial estimates for \teff\ were derived by ensuring no trend of Fe I line abundances with excitation potential (excitation equilibrium). \logg\ was adjusted until the same abundances were obtained from both Fe I and Fe II lines (ionization equilibrium). \vt\ was determined by ensuring no trends for Fe I abundances with reduced equivalent widths. The [Fe/H] values were determined from the mean of Fe I and Fe II lines after iterating to derive the LTE parameters for \teff, \logg, and \vt.

As discussed in \citet{frebel2013}, \citet{rpa3} and \citet{Ban2024ApJS}, the FR13$_{\rm corr}$ parameters are more reliable approximations and hence, we adopt the FR13$_{\rm corr}$ stellar parameters for abundance derivations in this study.

The corrected \teff\ values were determined using the empirical calibration from \citet{frebel2013}: 
\[
    T_{\mathrm{eff}}({\rm FR13}_{\rm corr}) = 0.9 \times T_{\mathrm{eff}}\text{(LTE)} + 670
\]

To evaluate the consistency of our adopted temperature scale, we derived photometric temperatures using Gaia magnitudes and the calibration presented in \citet{muc2021gaia}. This approach employs a polynomial relation between stellar color, metallicity, and the inverse temperature parameter $\theta = 5040/T_{\rm eff}$, specifically optimized for red giant stars. The relation includes cross-terms between color and metallicity, providing an accurate empirical framework for temperature estimation. Table~\ref{tab:temp} summarizes the comparison between the spectroscopic LTE temperatures, the spectroscopic temperatures corrected using the empirical relation of \citet{frebel2013}, and the Gaia-based photometric temperatures. Overall, we find that the Gaia-based photometric temperatures are systematically lower than the FR13$_{\rm corr}$ spectroscopic \teff\ by 33-195 K with the median difference being less than the associated uncertainty of 150 K. This level of discrepancy is well studied and discussed in detail by \citet{Mucboni2021}, who discuss systematic differences between spectroscopic and photometric temperatures in globular cluster stars. For consistency with our spectroscopic analysis and abundance determinations, we adopt the FR13$_{\rm corr}$  spectroscopic \teff\ throughout this work. The good overall agreement within uncertainties supports the robustness of the adopted temperatures.

\begin{figure}
\centering
\includegraphics[scale=0.4]{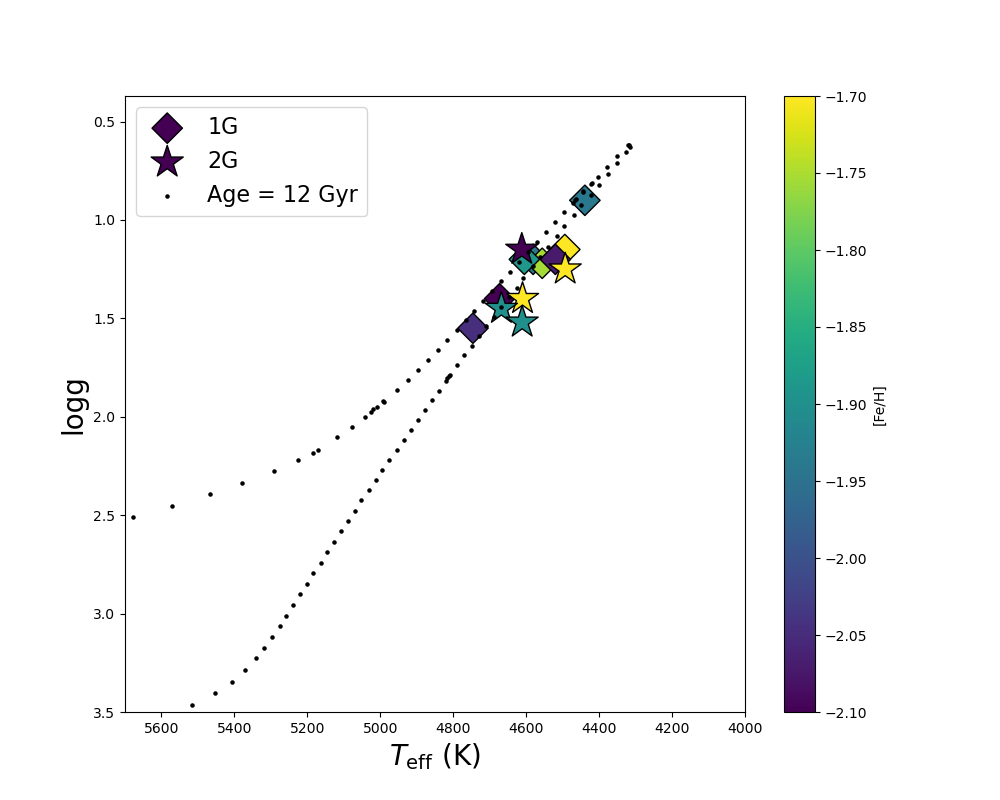}
\caption{Color-magnitude diagram showing the sample stars color-coded by metallicity, as indicated by the color bar. The stellar evolutionary track corresponds to age of 12 Gyr for a metallicity of [Fe/H] $= -2.0$.}
\label{cmd}
\end{figure}

After deriving the corrected \teff\ (denoted as \teff(FR13$_{\rm corr}$)), we re-derived our spectroscopic \logg, \vt, and [Fe/H]. Especially for the coolest stars, the FR13 correction produces significantly warmer temperatures ($\sim$ 200 K) that results in higher \logg\ and [Fe/H] for the target stars. As a note of caution, it is well established that NLTE effects can impact the ionization balance in metal-poor red giants, leading to a systematic underestimation of surface gravity when LTE is assumed \citep{bergemann2012,lind2012,rpa3}. This arises primarily due to overionization of Fe I levels, which leads to lower Fe I abundances relative to Fe II, thereby biasing the derived gravity lower in order to enforce equilibrium. Despite this, several studies have demonstrated that for typical globular cluster giants studied here at metallicities of [Fe/H] $\sim$ –1.98, NLTE corrections to surface gravity are on the order of 0.1–0.3 dex \citep{amarsi2016,rpa3}, and in many cases, the corrections lie within the uncertainties propagated from temperature and metallicity. In this work, we have chosen to retain LTE gravities for consistency with our abundance analysis, but we caution that the inferred log g values may be slightly underestimated, particularly for the most luminous giants in the sample. We emphasize, however, that the derived abundances of most elements, especially those based on neutral lines are only weakly sensitive to log g variations at this level. Our final corrected stellar parameters adopted for this study are listed in Table~\ref{tab:r_process_classification} and the corresponding color-magnitude diagram is shown in Figure \ref{cmd}. The isochrone is taken from \citet{Marigo2017} and corresponds to age of 12 Gyr, [$\alpha$/Fe] = +0.40 and [Fe/H]= $-$2.0. The stars are color-coded by metallicity and separated into first and second generation.

Our derived cluster metallicity of [Fe/H] = $-1.98$ is in excellent agreement with the optical spectroscopic estimate of [Fe/H] = $-1.91$ by \cite{mc92}, but $\sim$0.2 dex lower than [Fe/H] = $-1.71$ reported by \citet{baeza2022} based on NIR spectroscopic data. This analysis from GHOST supports a more metal-poor characterization of NGC~2298, consistent with earlier optical studies. Despite the 0.2 dex offset from \citet{baeza2022}, the trends and the nature of the [X/Fe] distribution—particularly for lighter elements such as Mg and Al—remain consistent, supporting the presence of abundance variations within the cluster.

\begin{figure*}[htbp]
    \centering
    \includegraphics[scale=0.5]{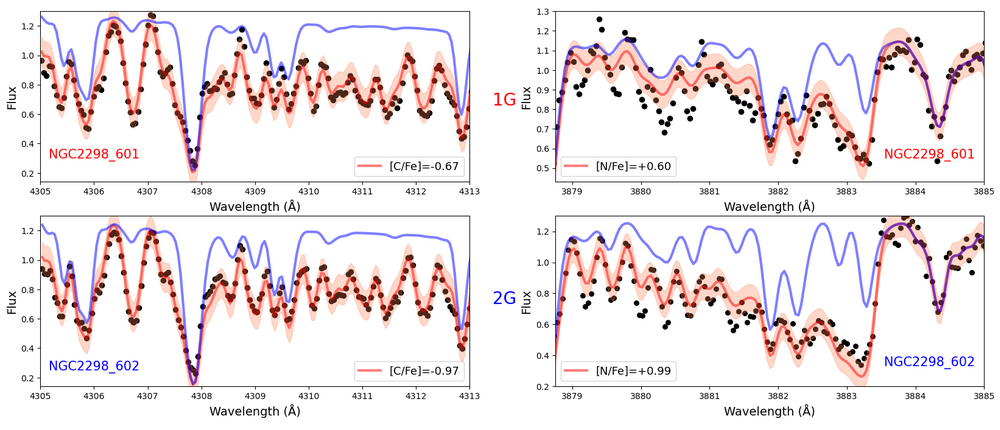}
    \includegraphics[scale=0.45]{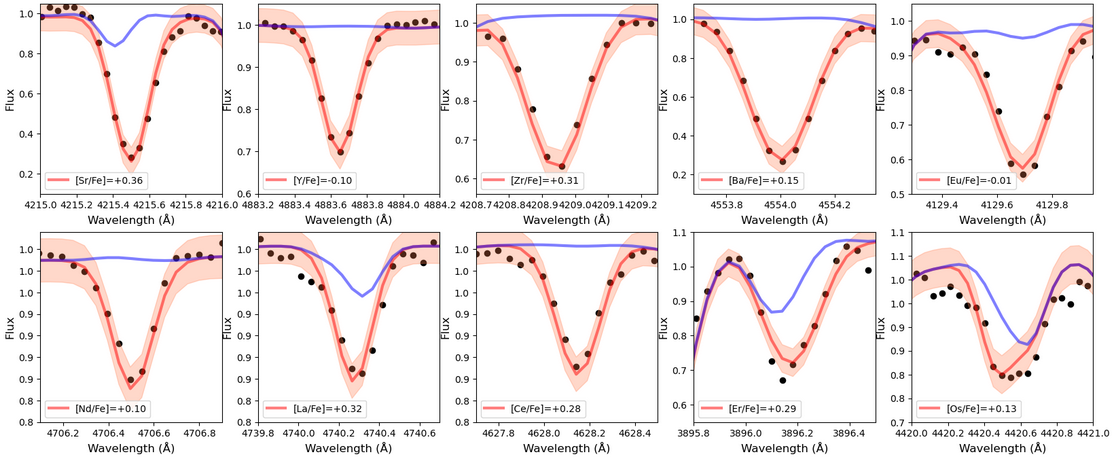}
    \caption{Top: Example spectral synthesis for the region of the molecular CH $G$-band and the molecular CN band to the left and right, respectively. The fits for a first generation (1G) star is shown in the top row and a second generation (2G) star is shown in the second row. The red line shows the best-fit synthetic spectrum to the data (black dots). The blue line corresponds to the absence of the respective element. The shaded region denotes the uncertainty in the measurement. Bottom:  Spectral synthesis for various neutron-capture elements, following the same color schemes.}
    \label{synthmol}
\end{figure*}

\begin{table*}
\centering
\caption{Stellar parameters and Classification of Program Stars}
	\label{tab:temp}
	\begin{tabular}{lcrrr}
		\hline
		Star name &Spectroscopic (LTE) \teff\ (K) &FR13$_{\rm corr}$ \teff\ (K) & Photometric \teff\ (K) \\
		\hline

NGC2298-11 &4448 &4673  &4520\\
NGC2298-12 &4345 &4581 &4495\\
NGC2298-13 &4339 &4575 &4380\\
NGC2298-14 &4377 &4610 & 4457\\
NGC2298-581 &4442 &4668 &4601 \\
NGC2298-582 &4247 &4493 & 4371 \\
NGC2298-591 &4317 &4556 &4491\\
NGC2298-592 &4187 &4439 &4316\\
NGC2298-601 &4528 &4746 &4612\\
NGC2298-602 &4379 &4611 &4430\\
NGC2298-611 &4372 &4605 &4481\\
NGC2298-612 &4249 &4494 &4461\\
NGC2298-981 &4278 &4520 &4438\\
		\hline
	\end{tabular}
\end{table*}

\section{Abundances}

Utilizing \texttt{MOOG} embedded in \texttt{SMHr}, we derive abundances and upper limits for the light elements, $\alpha$-elements, Fe-peak elements, and neutron-capture elements of all target stars. This includes 36 species from C to Th. We measured  equivalent widths (EWs) of the absorption lines across the entire optical spectral range, focusing on lines with EW $\leq 150$ m{\AA} and reduced equivalent widths (REWs) $\leq -4.5$ whenever feasible, as these lie on the linear portion of the curve of growth and are less sensitive to a given microturbulence \citep{Ban2018}. For all neutron-capture elements and some of the lighter elements, we used spectral synthesis to derive the abundances. The complete abundances of all the elements measured are given in Table 5.

The linelists, along with the isotopic and hyperfine structure data for relevant lines, including those of neutron-capture elements, were sourced from the R-Process Alliance (RPA) standard linelists with updated $\log gf$ values as detailed by \citet{roederer2018}. These linelists were generated using \texttt{linemake}\footnote{https://github.com/vmplacco/linemake} \citep{linemake_vini}. Solar photospheric abundances were adopted from \citet{asplund2009}. The linelist used in this study to derive the abundances for one star is provided in Table 4, the rest will be made available electronically.

\begin{table*}
\scriptsize
\centering
\caption{Stellar parameters and Classification of Program Stars}
\label{tab:r_process_classification}
\begin{tabular}{lcrrrrrrrrrrrrl}
\hline
Star name  & \teff (K) & \logg\ & [Fe/H] & \vt\ & [C/Fe] & $\Delta$[C/Fe]$^{a}$ & [N/Fe] & [Na/Fe] & [Mg/Fe] & [Al/Fe] & [Sr/Fe] & [Ba/Fe] & [Eu/Fe] & Pop. \\
\hline
NGC2298-11  & 4673 & 1.40 & $-$2.04 & 1.40 & $-$0.68 & 0.55 & 0.66 & 0.03  & 0.37 & 0.62 & 0.13 & $-$0.24 & 0.00  & 1G \\
NGC2298-12  & 4581 & 1.20 & $-$1.98 & 1.73 & $-$0.83 & 0.66 & 0.51 & $-$0.17 & 0.57 & 0.01 & 0.23 & $-$0.06 & $-$0.04 & 1G \\
NGC2298-591 & 4556 & 1.22 & $-$1.91 & 1.63 & $-$0.97 & 0.67 & 0.55 & $-$0.20 & 0.37 & 0.12 & 0.21 & $-$0.05 & $-$0.17 & 1G \\
NGC2298-592 & 4439 & 0.90 & $-$1.98 & 1.81 & $-$1.10 & 0.74 & 0.47 & $-$0.31 & 0.55 & $-$0.01 & 0.21 & 0.06  & $-$0.21 & 1G \\
NGC2298-601 & 4746 & 1.55 & $-$2.02 & 1.87 & $-$0.67 & 0.48 & 0.65 & 0.01  & 0.32 & 0.08 & 0.08 & $-$0.15 & 0.06  & 1G \\
NGC2298-611 & 4605 & 1.20 & $-$1.96 & 1.65 & $-$0.89 & 0.67 & 0.80 & 0.16  & 0.33 & 0.37 & 0.29 & $-$0.14 & $-$0.04 & 1G \\
NGC2298-612 & 4494 & 1.15 & $-$1.89 & 1.72 & $-$0.93 & 0.70 & 0.50 & $-$0.27 & 0.32 & $-$0.10 & 0.30 & 0.19  & 0.12  & 1G \\
NGC2298-981 & 4520 & 1.20 & $-$2.03 & 1.76 & $-$0.93 & 0.67 & 0.61 & $-$0.18 & 0.36 & 0.07 & 0.24 & 0.07  & $-$0.04 & 1G \\
\hline
Mean (1G)   & & &$-$1.97 & & $-$0.87 & & 0.59 & $-$0.11 & 0.39 & 0.14 & 0.21 & $-$0.04 & $-$0.04 & \\
\hline
NGC2298-13  & 4575 & 1.05 & $-$2.00 & 1.38 & $-$1.24 & 0.71 & 1.34 & 0.39  & 0.21 & 1.02 & 0.36 & 0.08  & 0.05  & 2G \\
NGC2298-14  & 4610 & 1.40 & $-$2.03 & 1.92 & $-$1.16 & 0.57 & 1.20 & 0.34  & 0.15 & 0.97 & 0.32 & $-$0.04 & $-$0.02 & 2G \\
NGC2298-581 & 4668 & 1.45 & $-$1.97 & 1.35 & $-$1.02 & 0.50 & 1.10 & 0.34  & 0.19 & 0.91 & 0.31 & $-$0.02 & $-$0.05 & 2G \\
NGC2298-582 & 4493 & 1.25 & $-$1.94 & 1.47 & $-$1.26 & 0.62 & 1.00 & 0.34  & 0.13 & 0.77 & 0.36 & 0.15  & $-$0.01 & 2G \\
NGC2298-602 & 4611 & 1.52 & $-$1.97 & 1.70 & $-$0.98 & 0.50 & 0.95 & 0.17  & 0.23 & 0.66 & 0.34 & 0.04  & 0.04  & 2G \\
\hline
Mean (2G)   &  &  & $-$1.98 & & $-$1.13 & & 1.11 & 0.31 & 0.18 & 0.86 & 0.33 & 0.04 & 0.0 & \\
\hline
\end{tabular}
\newline
$^a$ Indicates correction for evolutionary effects from \citet{placco2014}.
\end{table*}

\subsection{Non LTE Corrections}

Departures from local thermodynamic equilibrium (LTE) can introduce systematic errors in abundance determinations, especially in metal-poor giant stars where the radiation field departs significantly from the Planck function and collisional rates are low.
Among the odd Z elements, NLTE corrections are crucial for Al and Na. For Al, NLTE corrections were applied using the grid of \citet{nordlander2017}, which provides abundance corrections as a function of stellar parameters (effective temperature, surface gravity, metallicity, and line equivalent width). The magnitude of the correction increases with luminosity and decreasing surface gravity, varying from $\sim$+0.50\,dex at the base of the RGB to as high as +1.10\,dex near the RGB tip. Corrections were interpolated for each star in our sample based on its atmospheric parameters and applied to the LTE abundances. These corrected values are listed in Table~\ref{tab:r_process_classification}. NLTE corrections for Na were similarly adopted from the work of \citet{lind2011}, who provide detailed corrections for the Na~I D lines as well as other commonly used transitions. For our sample stars, corrections typically range between $-0.15$ and $-0.25$\,dex, with the exact value depending primarily on the stellar gravity and line strength. These corrections were applied uniformly, as Na is well known to be affected by strong over-recombination in metal-poor giants. For Mg, we evaluated NLTE effects based on the grid provided in \citet{Bergemann2017} using the online interface\footnote{https://nlte.mpia.de/}. In most cases, the NLTE corrections are modest, up to $\sim$+0.12\,dex, depending on the specific Mg~I lines used and the atmospheric parameters. However, these corrections are within the typical observational uncertainties for Mg abundances ($\sim$0.10--0.15\,dex), and thus were not applied to our measurements.

We also investigated NLTE effects for the iron-peak and neutron-capture elements. For many elements including Fe, the NLTE corrections tend to be small ($\lesssim$0.10\,dex) for the neutral species in cool giants \citep{lind2011}, while the ionized species are even less affected by NLTE. For certain elements such as Mn, large NLTE effects are known to affect the strong resonance triplet at 4030–4034\,\AA, particularly in metal-poor stars \citep{Bergemann2008Mn}. To avoid this complexity and its associated modeling uncertainties, we deliberately excluded the resonance triplet and other lines strongly affected by NLTE whenever possible from our analysis. A detailed analysis of the NLTE corrections for the Fe-peak elements is provided in \citet{bergemann2010,bergemann2014,Ezzeddine2016a,Mashonkina2023} and the references therein). For neutron-capture elements such as Sr, Ba and Eu, NLTE effects have been studied by \cite{bergemann2012,korotin2015nlte,Mashonkina2023}. In the stellar parameter range of our sample, corrections are predicted to be relatively low, which falls within the overall measurement uncertainties. For Sr and Ba the uncertainties are less than 0.10 dex while for Eu the corrections are slightly higher with corrections reaching 0.15 dex. As a result, we did not apply NLTE corrections to these species. However, we note that recent studies such as \citet{Mashonkina2023} emphasize the importance of NLTE modeling in specific contexts, and we caution that detailed line-by-line modeling might be required for extremely precise abundance work.

In summary, NLTE corrections were applied selectively—only where they are large enough to influence the interpretation of chemical abundances. For Al and Na, corrections are essential and significantly alter the inferred elemental trends and classification of the target stars. For other Fe-peak and neutron-capture elements, corrections are either small or the affected lines are excluded, ensuring the robustness of our abundance analysis. Importantly, given the relatively homogeneous stellar parameters within each population, the minor NLTE corrections would not affect the observed abundance dispersions or the conclusions drawn about intrinsic scatter and inter-population differences.

\subsection{Light elements}
Among the light elements, we detect and derive the abundances for C, N, Na, Mg, and Al. C and N abundances were derived based on spectral synthesis from the molecular bands of CH at 4313 \AA\ and CN at 3885 \AA. The carbon abundance derived from the CH band is employed to derive the nitrogen abundance, with iterative adjustments made to the N abundances to achieve the best fit to the observed CN band. We assumed the same ratios for $^{12}$C/$^{13}$C = 4.0 for the spectral analysis. An example of spectral synthesis of the molecular CH and CN bands for 1G and 2G stars are shown in the top and second panels of Figure \ref{synthmol} respectively. The Na abundances are obtained from the doublet D1 and D2 lines at 5896 and 5890 \AA\ while the Al abundances were derived from the strong absorption features 3944 and 3961 \AA, taking into account the blend with CH. For Mg abundances, we refrain from using the strong feature at 5183 \AA\ as it lies off the linear part of the curve of growth and may not be reliable for determining accurate abundances. The lines at 4703, 5172, 5528 and 5711 \AA\ were used for the determination of Mg abundances and deliver mostly consistent results based on the EW analysis. We examined the potential saturation of the strong Mg I lines blueward of 5711 \AA, including the Mg b triplet. In our abundance analysis, we primarily used weaker Mg lines such as Mg I at 5528 {\AA} when available, which are less affected by damping or saturation. For stars where Mg b lines were employed, we carefully inspected the profile fits and ensured they provided consistent abundances compared to other Mg lines. Any lines showing significant saturation effects were excluded from the final analysis.

\subsection{$\alpha$- and Fe-peak elements}

Among the $\alpha$-elements, we derive abundances for Si, Ca, Ti I, and Ti II along with Mg abundances which were discussed earlier. Several lines are detected for Ca and Ti with consistent line abundances throughout the optical spectra. For Si, we measure the strong transition at 3905 \AA\ (taking the C blend into account) for all the stars while the weaker transition at 4103 \AA\ is detected for some of the brighter targets.

We also measure abundances for Sc, V, Cr, Mn, Co, Ni, Cu, and Zn among the Fe-peak elements. The Fe abundances are based on the mean of the Fe I and Fe II lines in the spectra. A mean metallicity of [Fe/H]=$-$1.98 is obtained by computing the mean abundances or all the program stars in the cluster. We use equivalent widths for majority of the transitions while spectral analysis is employed whenever necessary to include any blends or hyperfine transitions. Multiple lines are used to derive the final abundances for each element and the complete results are provided in the Appendix.

\subsection{Neutron-capture elements}

We derive the abundances or meaningful upper limits of 16 neutron-capture elements for our sample stars. The elements include Sr, Y, Zr, Ba, La, Ce, Pr, Nd, Sm, Eu, Gd, Ho, Er, Os, and Th. Spectrum synthesis is employed to derive the abundances that include hyperfine structure or multiple isotopes. Isotopic ratios are taken from \citet{sneden2008}. An example of the spectral synthesis for several important n-capture elements is presented in the lower panels of Figure \ref{synthmol}. The abundances of all the n-capture elements are given in Table 3. The mean abundance of each element is adopted for elements with multiple lines in the spectra.

As a reference, for Eu, the abundances are derived from the strong line at 4129 \AA\ along with the weaker line at 4205 \AA. Similarly for Ba, the abundances are derived form the lines at 4554, 5853 and 6142 \AA.

\subsection{Abundance uncertainties}

The uncertainties in our abundance measurements stem from two primary sources: the SNR of the spectra, which affects the precision of the line fitting, and the uncertainties associated with the stellar parameters. To quantify the influence of SNR on our results, we apply the approach of \citet{Cayrel88}, as detailed in \citet{banli}, allowing us to estimate the related uncertainties. These uncertainties, typically under 0.1 dex, suggest minimal impact from SNR variations. Conversely, the uncertainties tied to stellar parameters are more systematic in nature and are evaluated following the procedures in \citet{ji2016b} and \citet{rpa3}. We account for potential deviations by considering changes of 150 K in effective temperature (\teff), 0.25 dex in surface gravity (log $g$), and 0.2 km s$^{-1}$ in microturbulence. Moreover, uncertainties in metallicity ([Fe/H]) are gauged by examining the dispersion in Fe I and Fe II line abundances. The total systematic uncertainty is then obtained by combining these individual contributions in quadrature. The uncertainties for each measurement for all the stars are provided with the abundances in Table 5.

\section{Discussion and Results} \label{results}

\subsection{Light element anti-correlation}

\begin{figure*}
\centering
\includegraphics[width=2.10\columnwidth]{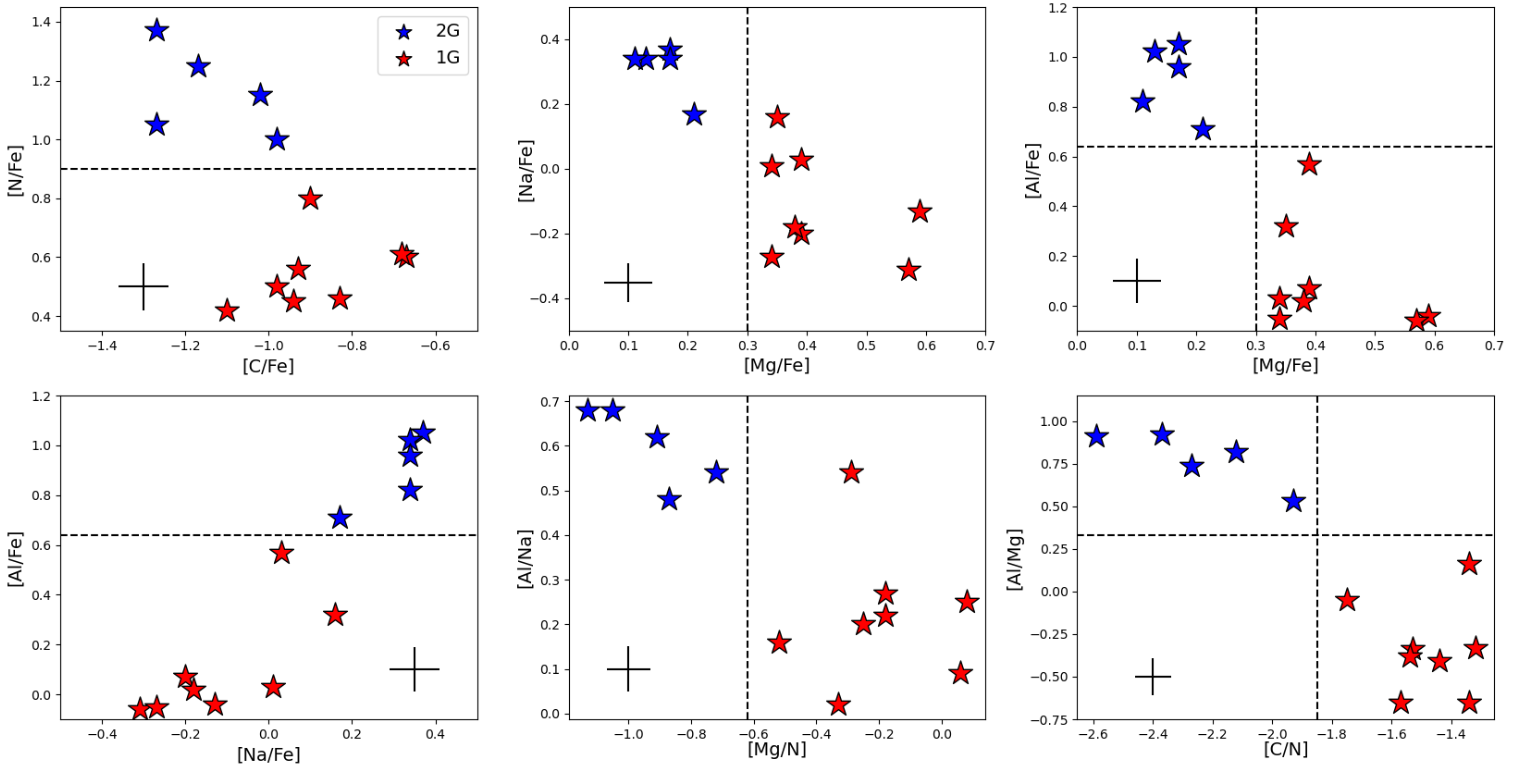}
\caption{Light element anti-correlations for 1G stars marked in red filled diamonds and 2G stars marked in blue filled diamonds. The dotted black line separates the 1G population from the 2G population.}
\label{anti}
\end{figure*}

We have classified our sample stars into two groups, first generation (1G) and second generation (2G) stars, based on the observed light element anti-correlations as demonstrated in Table \ref{tab:r_process_classification}. We could identify eight first generation stars and five second generation stars in NGC 2298, underscoring the strong presence of multiple stellar populations within this cluster. The results are shown in Figure \ref{anti}. 

We employ the CH-CN dichotomy to effectively separate the stars into these distinct groups as seen in the top left panel. 2G stars are found to have higher CN abundances. The same stars are also found to have higher [Na/Fe] and lower [Mg/Fe] as shown in the second panel of Figure \ref{anti}. The separation is further supported by the Mg-Al anticorrelation shown in the third panel that the stars follow in the same way. Specifically, the 2G stars exhibit high abundances of Na, Al and N, while showing low levels of Mg and C. The 1G stars on the contrary are found to exhibit lower abundances of Na, Al and N with comparatively higher levels of C and Mg. The same traits could be seen in the lower panels with the 2G stars and the 1G stars occupying distinct positions in the different abundance planes. The separation between the 1G and 2G stars is indicated by the black dotted line for each abundance plane.
This chemical signature aligns with patterns observed in other globular clusters, where similar abundance behaviors have been reported \citep{carretta2009b,gratton2012}, and is consistent with previous findings by \citet{bastian2017} and \citet{milone2022}, who extensively studied these traits in multiple stellar populations within metal-poor globular clusters. The mean abundances for the 1G and 2G stars are also provided in Table \ref{tab:r_process_classification}.

Among the 1G stars in our sample, we identify two stars (IDs 12 and 592) that exhibit significantly enhanced [Mg/Fe] ratios ($\sim +0.55$ dex), notably higher than the bulk of the 1G population, which shows a mean [Mg/Fe] of $\sim +0.35$ dex with low scatter. These enhancements are robust, as confirmed by visual inspection of the Mg\,\textsc{I} lines, which appear clean and well fit by the synthetic spectra. The elevated [Mg/Fe] values in these stars exceed the typical internal uncertainties ($\sim$0.05 dex) by more than 3$\sigma$. While globular cluster first-generation (1G) stars are generally expected to display a high degree of chemical homogeneity in light elements, recent high-precision spectroscopic surveys have revealed small abundance variations even among 1G stars \citep{gratton2019}. Such inhomogeneities may arise from early stochastic enrichment by core-collapse supernovae or localized chemical inhomogeneities in the proto-cluster gas. The two Mg-enhanced 1G stars in our sample may reflect such early inhomogeneities, or alternatively, they may represent a more primordial stellar population within the cluster that formed even earlier than the bulk of the 1G stars. Similar interpretations have been proposed for chemically distinct sub-populations within 1G stars in other clusters such as NGC~6752 \citep{carretta2017}. Notably, the Na and Al abundances of these two stars remain consistent with those of the other 1G stars, supporting their classification as first-generation members.

We note that a few first-generation stars exhibit [Na/Fe] values that slightly overlap with those of the second-generation population. However, as mentioned earlier our classification of multiple populations does not rely on a single abundance indicator. Instead, it is based on a combination of [C/Fe], [N/Fe], [Na/Fe], [Al/Fe], and [Mg/Fe], along with diagnostic correlations and anticorrelations (e.g., Na–Al and Mg–Al), ensuring a consistent and robust distinction. Similar minor overlaps in individual abundance planes have also been observed in other globular clusters with well-defined multiple populations, with notable examples such as M92 \citep{kirby23} and NGC 2808 \citep{Carretta-ngc2808-2015}.

We also note that we could not detect O which can add to the understanding of the nucleosynthetic processes influencing the different stellar populations. 

\subsection{$\alpha$- and Fe-peak elements}

\begin{figure*}
\centering
\includegraphics[width=2.05\columnwidth]{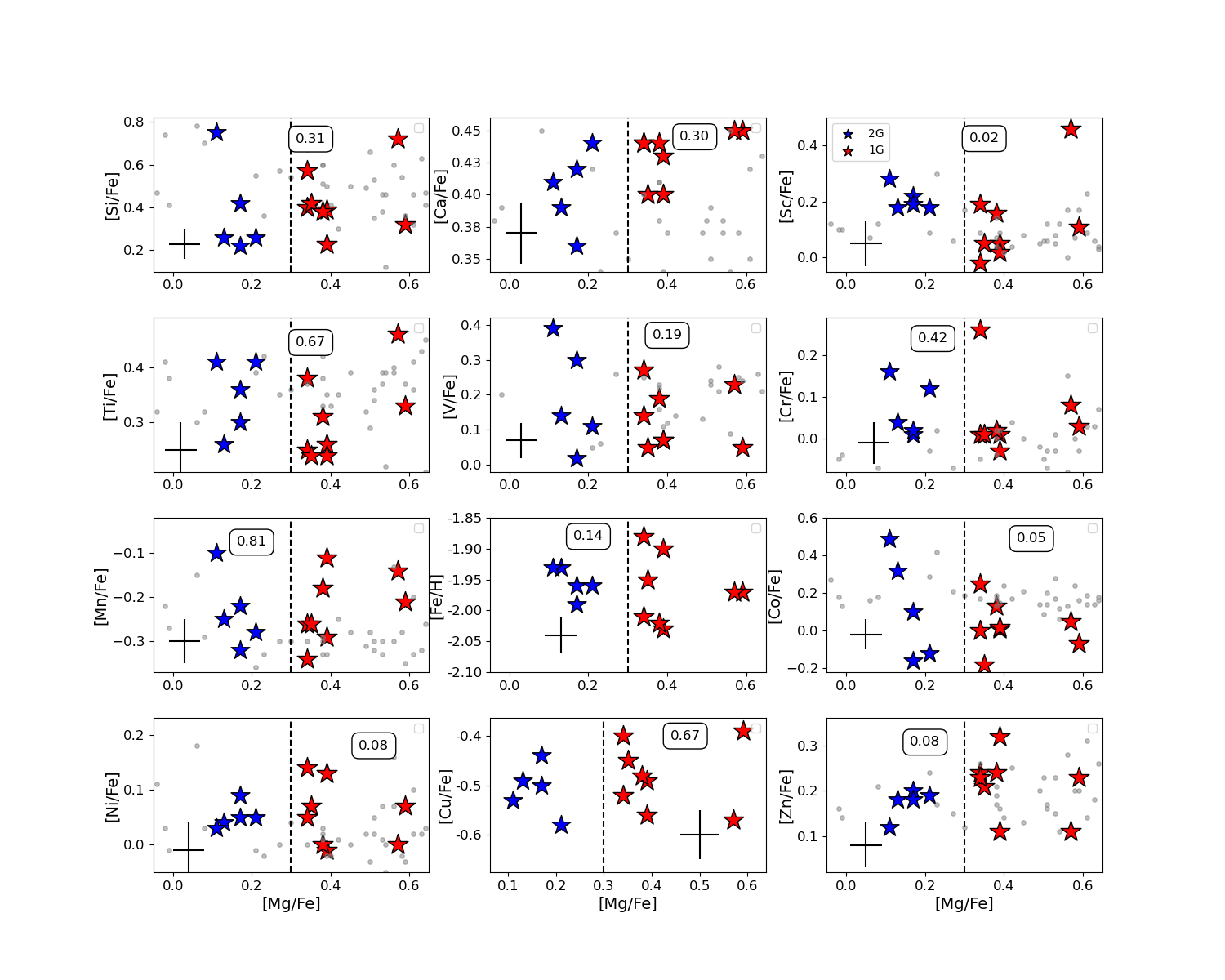}
\caption{Distribution of $\alpha$ and Fe-peak elements as a function of [Mg/Fe] for the 1G and 2G stars stars marked in red and blue filled diamonds, respectively. The p-values from the Brown-Forsythe test are labeled for each element in the respective panels. The grey dots correspond to abundances in M92 taken from \citet{kirby23}.}
\label{Trends}
\end{figure*}

Following the light elements, we analyze the trends of $\alpha$ and Fe-peak elements with Mg in both 1G and 2G stars, as illustrated in Figure \ref{Trends}. Using the same color scheme as in Figure \ref{anti}, we find a similar dispersion for the majority of the elements in the 1G and 2G stars. Notably, both the 1G and 2G stars exhibit a remarkably small dispersion in calcium ($<$ 0.10 dex). This relatively small dispersion in Ca abundances suggests a homogeneity in the gas mixing process, which may point to a common progenitor supernovae.

However, we observe that the dispersion is significantly lower for Sc, Ni and Zn in the 2G stars compared to the 1G stars. The dispersion for the 1G and 2G stars is compared by employing the Brown-Forsythe test to assess the equality of variances \citep{brownfs}.  The p-value represents the probability of observing a difference in dispersions similar to what is observed, assuming that there is no true difference between the groups. The resulting p-value is labeled for each element in the corresponding panel in Figure \ref{Trends}. Any p-value lower than 0.10 indicates intrinsic difference in the dispersion between the two populations while values higher than 0.10 indicates no such inequalities \citep{wilcox2017}. The minimal dispersion of Sc among 2G stars with a p-value of 0.02 corresponding to dispersions greater than 2$\sigma$ suggests a significant contribution from a single mass progenitor to the gas from which 2G stars form, possibly indicating a specific phase of supernova pre-enrichment. However, the large dispersion among 1G stars may indicate progenitors with a wide range of masses, whose yields were well mixed into the gas prior to the formation of 2G stars. In contrast, the lack of dispersion in Ni and Zn among 2G stars suggests dominant contributions from single progenitors. Our findings align with the results from previous studies, albeit on dwarf galaxies, which  suggested that the chemical signatures of stars that are formed later in time are shaped by a narrower range of progenitor masses compared to the earliest populations \citep{venn2004,kobayashi2006}. This also indicates that NGC 2298 did not form from a perfectly homogeneous gas cloud but rather from a gas that was variably enriched by multiple nucleosynthetic events (e.g., multiple supernovae with different characteristics). The first-generation stars inherited these variations directly which might have been diluted or homogenized by the time second generation stars were formed. The early environment could also have experienced inhomogeneous mixing of the ISM, resulting in local pockets with different elemental abundances. This is consistent with the higher disparities in the degree of variations seen in Sc and the more moderate but still significant spread in Ni and Zn.

We also observe a notably large dispersion of similar scale in Ti, Mn and Cu abundances for both the 1G and 2G stars as signified by the high p-values. The larger dispersion in these elements suggests contributions from a more varied set of nucleosynthetic processes or progenitor stars with different masses and explosion energies. For example, Cu is known to be produced in multiple sites, including asymptotic giant branch (AGB) stars through weak s-process nucleosynthesis, which could account for the observed spread \citep{bisterzo}.

\subsection{Nucleosynthesis patterns}

Our study highlights significant differences in the abundances of light elements such as C, N, Na, Mg, and Al between the 1G and 2G stars, indicated by red and blue colors, respectively in the top panel of Figure \ref{chart}. The odd-even pattern observed for odd-Z elements reflects typical nucleosynthetic processes associated with supernovae, particularly those near the iron peak, highlighting the diversity of supernova progenitors contributing to the chemical enrichment of the cluster. The abundance patterns of Fe-peak and $\alpha$-elements observed in our study are consistent with the expected yields from core-collapse supernovae (CCSNe), which dominate early globular cluster formation \citep{carretta2010x}. The small dispersions in $\alpha$-elements, such as Ca, across both stellar generations suggest that these elements were produced in a well-mixed interstellar medium prior to the formation of 2G stars. However, more than 2-sigma spread for Sc and moderate spreads of around 1.67-sigma for Ni and Zn, compared to other iron-peak elements, may indicate localized inhomogeneities in the interstellar medium or distinct nucleosynthetic pathways that were more prominent during the cluster's early formation stages. Overall, despite variations in dispersion between the 1G and 2G stars for few of the elements, the distribution of lighter elements up to Zn displays typical Type II supernovae signatures.

\subsection{$r$-process elements}

\begin{figure*}
\centering
\includegraphics[width=1.90\columnwidth]{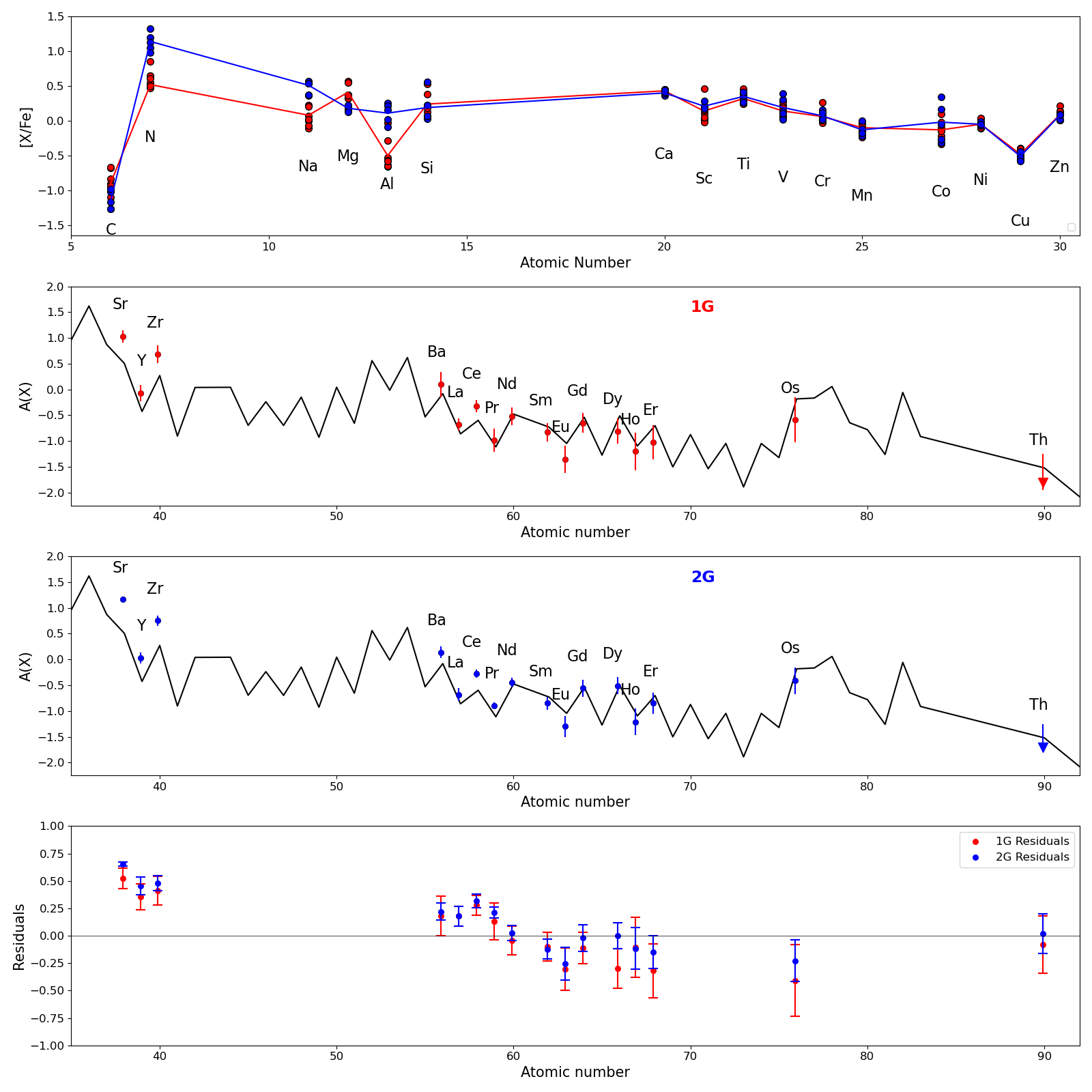}
\caption{Top: Significant differences exist in the abundances of light elements (C to Zn) between 1G stars (red) and 2G stars (blue). The next two panels present the abundances of 16 neutron-capture elements averaged for the 1G and 2G stars marked in red and blue respectively. The dispersion of the abundances within the 1G and 2G stars are marked by the vertical lines. The solar abundances are taken from \citet{Arlandini1999} and the scaled solar $r$-process pattern is normalized to the weighted average of the lanthanides. The residuals of the abundances of each element for the 1G and 2G stars are shown in the bottom panel.}
\label{chart}
\end{figure*}

Among the elements beyond Zn, we successfully detect and measure the abundances for 16 neutron-capture elements. They all exhibit the $r$-process abundance pattern, as shown in the middle panels of Figure \ref{chart}, where the abundances are averaged for the 1G and 2G stars and the respective dispersions are shown by vertical lines. The scaled solar r-process pattern \citep{sneden2008} is normalized to the weighted average of the lanthanides of the 1G and 2G stars, shown in red and blue, respectively.  
The residuals of the deviations from the scaled solar pattern are shown in the bottom panel in the respective colors.

The dispersion of the measured abundances within the 1G and 2G stars are also marked by vertical lines for each element in the respective colors. We find a larger dispersion for the heavy $r$-process elements compared the light $r$-process elements which has also been seen in other studies (e.g., \citealt{ji2016b}). Another intriguing aspect is the lower-than-expected Eu abundances \citep{holmbeck2018} for the stars of NGC 2298. This deviation suggests that there might be variations in the nature of the $r$-process contributions to Eu in this cluster's environment, or that the s-process could have played a role in the formation of other elements. Further investigation is needed to clarify these contributions. As seen in the bottom panel, the mean n-capture abundance patterns are statistically indistinguishable between the 1G and 2G populations.  This arises from the fact that the differences shown here do not extend beyond their mutual 1-$\sigma$ uncertainties.  However, the dispersion about the mean of the abundances in some elements remains statistically meaningful.

\subsection{Intrinsic differences between first and second generation stars}

To assess whether the observed dispersion in elemental abundances reflects genuine star-to-star variations rather than observational uncertainties, we estimated the intrinsic scatter ($\sigma_{\rm int}$) for each element in both the first- and second-generation (1G and 2G) stars using a maximum likelihood approach following the formalism of \citet{Walker2006}. For each population, we assume the observed abundance of a given element, $x_i$, for star $i$ is drawn from a Gaussian distribution with a mean abundance $\mu$ and a variance that is the sum of the measurement error $\sigma_i^2$ and an intrinsic dispersion $\sigma_{\rm int}^2$. The likelihood function is given by:

\begin{equation}
\mathcal{L} = \prod_{i=1}^{N} \frac{1}{\sqrt{2\pi (\sigma_i^2 + \sigma_{\rm int}^2)}} \exp\left( -\frac{(x_i - \mu)^2}{2(\sigma_i^2 + \sigma_{\rm int}^2)} \right).
\end{equation}

We maximize this likelihood with respect to both $\mu$ and $\sigma_{\rm int}$ to determine the most probable values. The results are summarized in Table~\ref{tab:intrinsic_scatter}.

Interestingly, we find that 1G stars show relatively higher intrinsic scatter in several elements, notably in [Sc/Fe] and the neutron-capture elements [Ba/Fe] and [Eu/Fe], where $\sigma_{\rm int} \sim 0.08$--$0.12$ dex. In contrast, 2G stars appear chemically more homogeneous in these elements, but exhibit a significant dispersion in [Co/Fe] ($\sigma_{\rm int} = 0.24$ dex). The broader spread in neutron-capture elements in 1G stars could reflect stochastic early enrichment from a small number of asymptotic giant branch or neutron star merger events, as has been noted in previous studies of globular clusters \citep{kirby23}. The enhanced Co spread in 2G stars may hint at variations in nucleosynthetic sources associated with the cluster's second generation, potentially linked to early supernova ejecta mixing. Overall, the derived intrinsic dispersions reinforce the presence of subtle but measurable chemical inhomogeneities within NGC 2298 that are generation-dependent.

\subsection{Light and heavy $r$-process}

\begin{figure*}
\centering
\includegraphics[width=2.30\columnwidth]{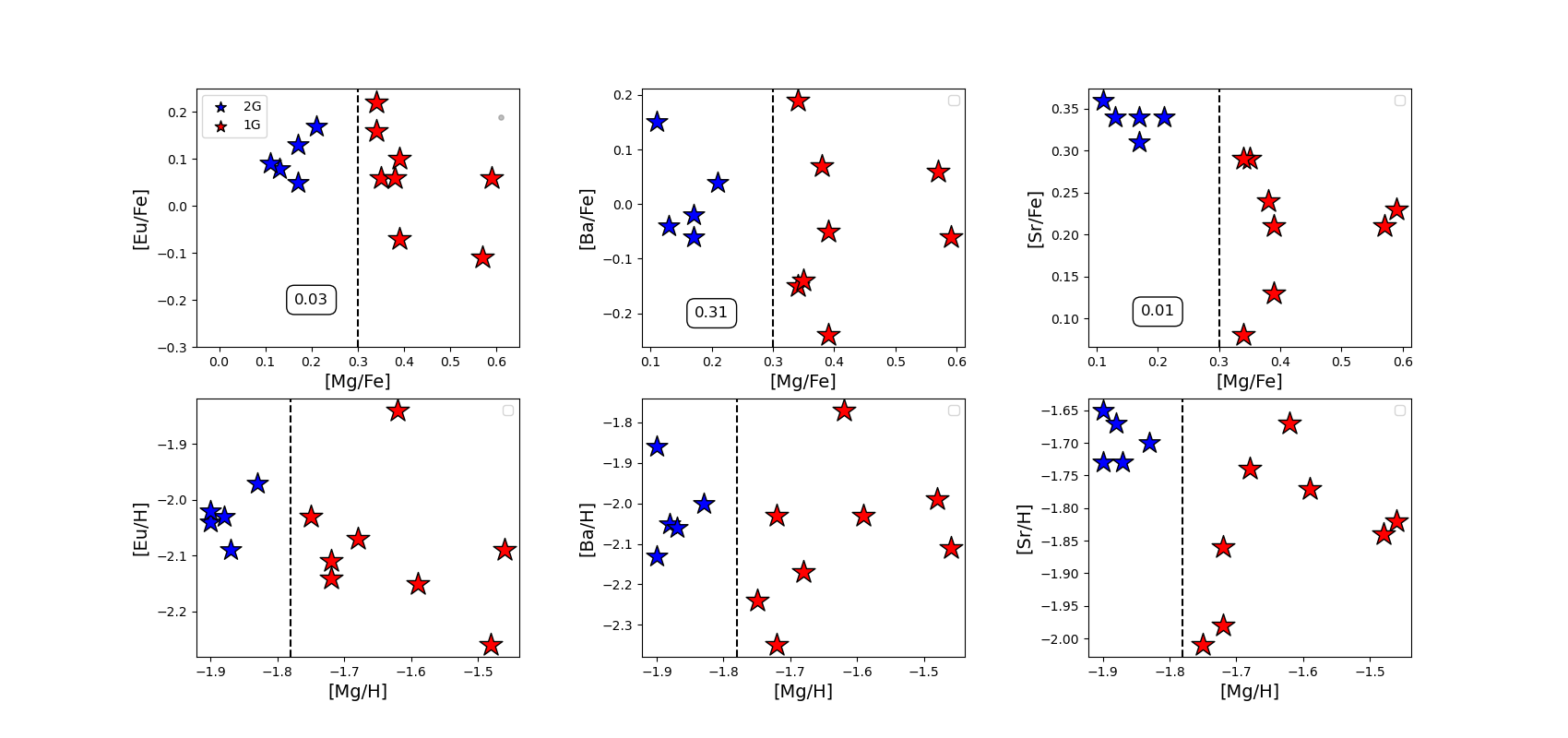}
\caption{Variations in Sr, Ba, and Eu abundances as a function of Mg. The top row presents the [X/Fe] ratios of Sr, Ba, and Eu plotted against [Mg/Fe]. The second row removes metallicity dependence by displaying [X/H] versus [Mg/H]. The bottom panel highlights the trends of Sr, representing a lighter r-process element, in comparison to the heavier r-process elements Ba and Eu.}
\label{srbaeu}
\end{figure*}

\begin{figure*}
\centering
\includegraphics[width=2.20\columnwidth]{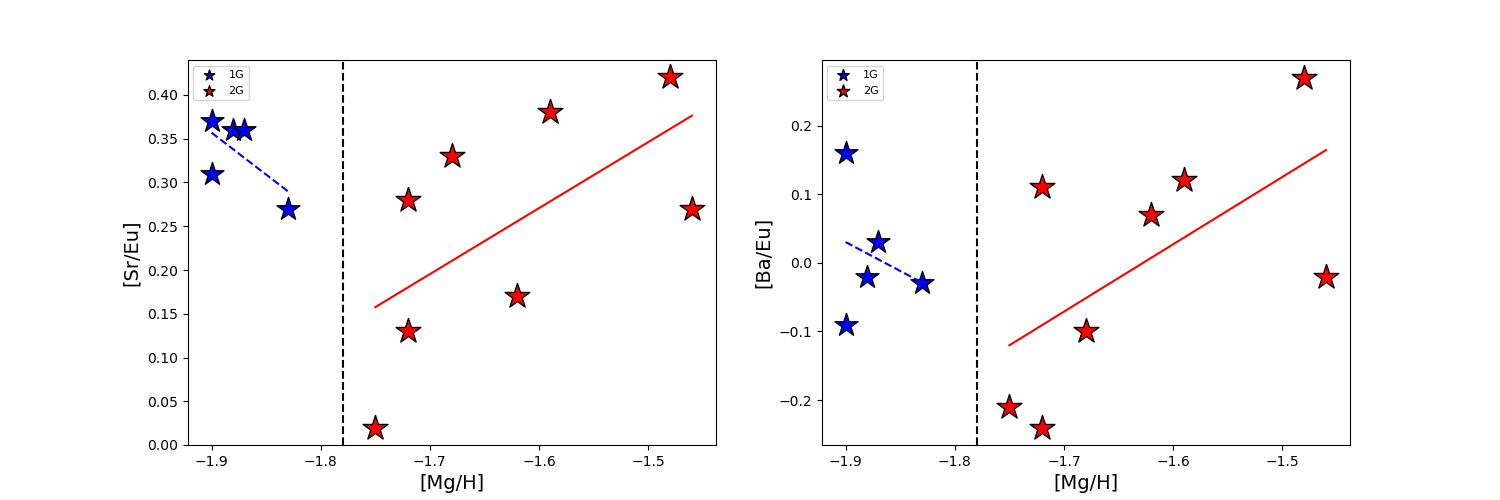}
\caption{Linear fits to the abundance ratios for the lighter and heavier $r$-process element [Sr/Eu] (left panel) and [Ba/Eu] (right panel) as a function of [Mg/H] for the 1G (blue) and 2G (red) stellar populations. The lines represent linear least-squares fits. The corresponding slopes and $p$-values are:
    \textbf{Left:} 1G [Sr/Eu] slope = $-0.958$, $p = 0.2399$; 2G [Sr/Eu] slope = $+0.754$, $p = 0.1004$.
    \textbf{Right:} 1G [Ba/Eu] slope = $-0.843$, $p = 0.6748$; 2G [Ba/Eu] slope = $+0.981$, $p = 0.1019$.
    These results suggest a mild, statistically non-significant trend, particularly in the 2G population.}
\label{srbaeu2}
\end{figure*}

In our analysis of the neutron-capture elements within the globular cluster NGC 2298, we also examined variations in Sr, Ba, and Eu abundances as a function of Mg. In the top row of Figure \ref{srbaeu}, we plotted the [X/Fe] ratios of Sr, Ba, and Eu against [Mg/Fe]. In the bottom row, we removed the dependence on metallicity by plotting [X/H] vs [Mg/H]. The results indicate that 1G stars exhibit a substantially higher dispersion in Sr and Eu abundances compared to 2G stars with p-values 0.01 and 0.03 corresponding to variations higher than 2 $\sigma$ levels. Specifically, the spread in Eu abundance is at a level of 0.1 dex for 2G stars but exceeds 0.5 dex for 1G stars. The p-values for the dispersion are printed in the top panels for Sr, Ba, and Eu. The 2-$\sigma$ variation in Sr among the 1G and 2G stars could suggest that the $r$-process contribution to the first-generation stars varied significantly, possibly indicating a heterogeneous enrichment by different $r$-process events, like neutron star mergers or rare supernova types with strong $r$-process contributions.  Similar observations for Eu abundance suggests that there was a significant stochastic element to the enrichment by $r$-process events. This could imply that first-generation stars formed from gas that was inhomogeneously enriched by a few localized $r$-process events. Such variations might also point to a scenario where $r$-process enrichment events were few and far between, and the gas from which the first-generation stars formed was not well-mixed, leading to significant local variations in Eu abundance. When we remove the dependence on metallicity, the trends for Sr, Ba, and Eu shown in the lower panels remain consistent with those observed in the top panels which is expected given the low dispersion in metallicity. This consistency highlights the robustness of our findings and suggests that the observed dispersions are intrinsic to the stellar populations rather than being artifacts of varying metallicity.

In Figure \ref{srbaeu2}, we analyze the trends of Sr as a lighter r-process element in comparison to the heavier r-process elements Ba and Eu \citep{tsuji1,Bandyo2020,banchemo} . On the left panel, a slight increasing trend of [Sr/Eu] with rising [Mg/H] was noted for the 1G stars. A similar, albeit modest, trend was observed for [Ba/Eu] in the right panel. While these trends may hint at a potential correlation between the activity of $r$-process nucleosynthesis responsible for producing lighter elements such as Sr, as well as Ba and regions with higher [Mg/H], the limited variation in [Mg/H] ($\sim 0.3$ dex) suggests that this finding should be interpreted with caution. The resulting slopes for 1G stars are -0.96 for [Sr/Eu] and -0.84 for [Ba/Eu], while the 2G slopes are +0.75 and +0.98 respectively. Although these trends are not statistically significant at the 95\% confidence levels (all $p$-values $> 0.10$), the sign and magnitude of the slopes suggest potential opposite behaviors between 1G and 2G stars: a weak decreasing trend for 1G stars and a weak increasing trend for 2G stars. This may hint at subtle differences in enrichment pathways between the generations, particularly for n-capture elements relative to  Mg, an $\alpha$ element.  Further studies with a bigger sample size is needed to confirm or refute the robustness of these premiliminary trends. This is particularly intriguing for the heavy neutron-capture elements, possibly pointing to varying contributions from the $r$-process or other enrichment sources between stellar generations. This finding supports the hypothesis that multiple r-process events, potentially from different types of progenitors, contributed to the chemical enrichment of the cluster \citep{truran2002}. 

We do not find signatures of an actinide boost based on the measurement of the upper limits for Th in this study. The actinide boost serves as a critical diagnostic in globular clusters, offering insights into the extent of $r$-process nucleosynthesis and the enrichment history of these systems. The upper limits for $\log \epsilon (\text{Th/Eu})$ values for both 1G and 2G stars are approximately $< -0.45$, despite the low Eu abundances. This value places them within the category of actinide normal stars as defined by \citet{holmbeck2018}. For comparison, the canonical actinide boost star CS 31082-001 \citep{Ernandes2023} exhibits $\log \epsilon (\text{Th/Eu}) = -0.10$, while the canonical actinide normal star CS 22892-052 \citep{Sneden2003} shows $\log \epsilon (\text{Th/Eu}) = -0.62$. Despite detecting several main $r$-process elements, the likely absence of an actinide boost suggests a limited or distinct contribution from the most extreme $r$-process events, such as neutron star mergers, in the early chemical evolution of NGC 2298. S-process is unlikely to have dominated the enrichment process as indicated by low C abundance, no detection of Li and no detection of Pb abundances at 4057 \AA. 

\begin{table}[h!]
\centering
\caption{Mean abundances and intrinsic scatter ($\sigma_{\rm int}$) for [X/Fe] in 1G and 2G stars in NGC 2298, derived using the maximum likelihood method.}
\begin{tabular}{lcc|cc}
\hline
Element & $\langle$[X/Fe]$\rangle$ (1G) & $\sigma_{\rm int}$ (1G) & $\langle$[X/Fe]$\rangle$ (2G) & $\sigma_{\rm int}$ (2G) \\
\hline
Si & 0.23   &0.12  & 0.18   &0.17   \\
Ca & 0.43   &0.0  & 0.41   & 0.0   \\
Sc & 0.12   &0.12   & 0.21   &0.0    \\
Ti & 0.30   & 0.04 & 0.34   & 0.0   \\
V & 0.13   & 0.04  & 0.19   & 0.11   \\
Cr & 0.05   &0.07  &0.07    &0.01    \\
Mn &-0.12    &0.0   &-0.14    & 0.04   \\
Co & -0.12   &0.07  & 0.02   & 0.24   \\
Ni &-0.04    &0.0   & -0.04   & 0.0   \\
Cu & -0.48   &0.0  & -0.40   & 0.0   \\
Zn &0.11    &0.02   & 0.07   & 0.0    \\
Sr & 0.21   &0.03  & 0.33   & 0.0   \\
Ba & -0.04   &0.12   & 0.04   & 0.03    \\
Eu & -0.03   & 0.08  & 0.01   & 0.0   \\
\hline
\end{tabular}
\label{tab:intrinsic_scatter}
\end{table}

\section{Conclusion} \label{conc}

Our study identified eight first-generation stars and five second-generation stars in NGC 2298 using the GHOST spectrograph, underscoring the presence of multiple stellar populations within this globular cluster. These findings significantly contribute to the growing body of evidence supporting the intricate and varied chemical evolution of globular clusters, specifically highlighting the unique characteristics of NGC 2298. By classifying stars into first and second generations based on light element abundance anticorrelations, we confirmed the presence of characteristic chemical anomalies within the cluster.

While the overall metallicity among cluster members remains uniform, the observed variations in specific $\alpha$- and Fe-peak elements, particularly Sc, Ni, and Zn, suggest generation-dependent nucleosynthesis processes or inhomogeneous mixing in the early cluster environment. The pronounced dispersion of neutron-capture elements, especially Sr and Eu with variations greater than 2$\sigma$ among the 1G and 2G stars, further underscores the complexity of the enrichment processes. We also confirm an intrinsic scatter beyond observational uncertainties for several elements using a maximum likelihood approach among stars from different generations. The observed potentially opposite trends in [Sr/Eu] and [Ba/Eu] ratios as a function of [Mg/H] for 1 G and 2G stars provides additional insights. The variations in Sc, Sr, and Eu abundances among first-generation stars point to a complex early chemical enrichment process, involving contributions from multiple nucleosynthetic sources. The significant scatter indicates inhomogeneous enrichment, likely resulting from rapid formation from gas that was unevenly mixed and influenced by localized supernovae and neutron-capture events.

These findings suggest a dynamic cluster formation environment shaped by contributions from a diverse array of massive stars, and rare $r$-process events. Our detection of the universal $r$-process pattern with significant dispersion, particularly in the main $r$-process elements, underscores the diverse origins of these elements in globular clusters. The absence of an actinide boost points to a consistent but complex chemical enrichment history. Overall, these results contribute to a more nuanced understanding of the formation and chemical evolution of metal-poor globular clusters and their role in the early history of the Milky Way.

\acknowledgements
A.B acknowledges the efforts of Prasanta Kumar Nayak and discussions on developing the methods of target selection discussed in this study.
R.E. acknowledges support from NSF grant AST 2206263 and NASA Astrophysics Theory Program grant 80NSSC24K0899. 
A.F. acknowledges support from  NSF grant AST-2307436.
The work of V.M.P. is supported by NOIRLab, which is managed by the Association of Universities for Research in Astronomy (AURA) under a cooperative agreement with the U.S. National Science Foundation. D.A. acknowledges financial support from the Spanish Ministry of Science and Innovation (MICINN) under the 2021 Ram\'on y Cajal program MICINN RYC2021‐032609. I.U.R.\ acknowledges support from NSF grant AST~2205847 and NASA Astrophysics Data Analysis Program grant 80NSSC21K0627.

\bibliography{sample631}{}

\begin{thebibliography}{}
\expandafter\ifx\csname natexlab\endcsname\relax\def\natexlab#1{#1}\fi
\providecommand{\url}[1]{\href{#1}{#1}}
\providecommand{\dodoi}[1]{doi:~\href{http://doi.org/#1}{\nolinkurl{#1}}}
\providecommand{\doeprint}[1]{\href{http://ascl.net/#1}{\nolinkurl{http://ascl.net/#1}}}
\providecommand{\doarXiv}[1]{\href{https://arxiv.org/abs/#1}{\nolinkurl{https://arxiv.org/abs/#1}}}

\bibitem[{{Amarsi} {et~al.}(2016){Amarsi}, {Lind}, {Asplund}, {Barklem}, \& {Collet}}]{amarsi2016}
{Amarsi}, A.~M., {Lind}, K., {Asplund}, M., {Barklem}, P.~S., \& {Collet}, R. 2016, \mnras, 463, 1518, \dodoi{10.1093/mnras/stw2077}

\bibitem[{Arlandini {et~al.}(1999)Arlandini, Käppeler, Wisshak, Gallino, Lugaro, Busso, \& Straniero}]{Arlandini1999}
Arlandini, C., Käppeler, F., Wisshak, K., {et~al.} 1999, The Astrophysical Journal, 525, 886, \dodoi{10.1086/307938}

\bibitem[{{Asplund} {et~al.}(2009){Asplund}, {Grevesse}, {Sauval}, \& {Scott}}]{asplund2009}
{Asplund}, M., {Grevesse}, N., {Sauval}, A.~J., \& {Scott}, P. 2009, \araa, 47, 481, \dodoi{10.1146/annurev.astro.46.060407.145222}

\bibitem[{{Baeza} {et~al.}(2022){Baeza}, {Fern{\'a}ndez-Trincado}, {Villanova}, {Geisler}, {Minniti}, {Garro}, {Barbuy}, {Beers}, \& {Lane}}]{baeza2022}
{Baeza}, I., {Fern{\'a}ndez-Trincado}, J.~G., {Villanova}, S., {et~al.} 2022, \aap, 662, A47, \dodoi{10.1051/0004-6361/202243475}

\bibitem[{{Bailin} \& {Harris}(2009)}]{bailin2009}
{Bailin}, J., \& {Harris}, W.~E. 2009, \apj, 695, 1082, \dodoi{10.1088/0004-637X/695/2/1082}

\bibitem[{{Balbinot} {et~al.}(2011){Balbinot}, {Santiago}, {da Costa}, {Makler}, \& {Maia}}]{bal2011}
{Balbinot}, E., {Santiago}, B.~X., {da Costa}, L.~N., {Makler}, M., \& {Maia}, M. A.~G. 2011, \mnras, 416, 393, \dodoi{10.1111/j.1365-2966.2011.19044.x}

\bibitem[{Bandyopadhyay {et~al.}(2024)Bandyopadhyay, Beers, Ezzeddine, Sivarani, Nayak, Pandey, Saraf, \& Susmitha}]{banchemo}
Bandyopadhyay, A., Beers, T.~C., Ezzeddine, R., {et~al.} 2024, Monthly Notices of the Royal Astronomical Society, 529, 2191, \dodoi{10.1093/mnras/stae613}

\bibitem[{Bandyopadhyay {et~al.}(2020{\natexlab{a}})Bandyopadhyay, Sivarani, \& Beers}]{Bandyo2020}
Bandyopadhyay, A., Sivarani, T., \& Beers, T.~C. 2020{\natexlab{a}}, The Astrophysical Journal, 899, 22, \dodoi{10.3847/1538-4357/ab9c9d}

\bibitem[{Bandyopadhyay {et~al.}(2022)Bandyopadhyay, Sivarani, Beers, Susmitha, Nayak, \& Pandey}]{banli}
Bandyopadhyay, A., Sivarani, T., Beers, T.~C., {et~al.} 2022, The Astrophysical Journal, 937, 52, \dodoi{10.3847/1538-4357/ac8b0f}

\bibitem[{Bandyopadhyay {et~al.}(2018)Bandyopadhyay, Sivarani, Susmitha, Beers, Giridhar, Surya, \& Masseron}]{Ban2018}
Bandyopadhyay, A., Sivarani, T., Susmitha, A., {et~al.} 2018, The Astrophysical Journal, 859, 114, \dodoi{10.3847/1538-4357/aabe80}

\bibitem[{Bandyopadhyay {et~al.}(2020{\natexlab{b}})Bandyopadhyay, Thirupathi, Beers, \& Susmitha}]{bangc}
Bandyopadhyay, A., Thirupathi, S., Beers, T.~C., \& Susmitha, A. 2020{\natexlab{b}}, Monthly Notices of the Royal Astronomical Society, 494, 36, \dodoi{10.1093/mnras/staa671}

\bibitem[{{Bandyopadhyay} {et~al.}(2024){Bandyopadhyay}, {Ezzeddine}, {Allende Prieto}, {Aria}, {Shah}, {Beers}, {Frebel}, {Hansen}, {Holmbeck}, {Placco}, {Roederer}, \& {Sakari}}]{Ban2024ApJS}
{Bandyopadhyay}, A., {Ezzeddine}, R., {Allende Prieto}, C., {et~al.} 2024, \apjs, 274, 39, \dodoi{10.3847/1538-4365/ad6f0f}

\bibitem[{{Bastian} \& {Lardo}(2018)}]{bastian2017}
{Bastian}, N., \& {Lardo}, C. 2018, \araa, 56, 83, \dodoi{10.1146/annurev-astro-081817-051839}

\bibitem[{{Beasley}(2020)}]{beasley2020}
{Beasley}, M.~A. 2020, in Reviews in Frontiers of Modern Astrophysics; From Space Debris to Cosmology, 245--277, \dodoi{10.1007/978-3-030-38509-5_9}

\bibitem[{{Bekki}(2017)}]{bekki17}
{Bekki}, K. 2017, \mnras, 467, 1857, \dodoi{10.1093/mnras/stx110}

\bibitem[{{Bekki}(2019)}]{Bekki19}
---. 2019, \mnras, 486, 2570, \dodoi{10.1093/mnras/stz999}

\bibitem[{Bekki {et~al.}(2007)Bekki, Yahagi, \& Forbes}]{Bekki2007}
Bekki, K., Yahagi, H., \& Forbes, D.~A. 2007, Monthly Notices of the Royal Astronomical Society, 377, 215, \dodoi{10.1111/j.1365-2966.2007.11588.x}

\bibitem[{{Bergemann}(2014)}]{bergemann2014}
{Bergemann}, M. 2014, ArXiv e-prints

\bibitem[{{Bergemann} \& {Cescutti}(2010)}]{bergemann2010}
{Bergemann}, M., \& {Cescutti}, G. 2010, \aap, 522, A9, \dodoi{10.1051/0004-6361/201014250}

\bibitem[{{Bergemann} {et~al.}(2017){Bergemann}, {Collet}, {Amarsi}, {Kovalev}, {Ruchti}, \& {Magic}}]{Bergemann2017}
{Bergemann}, M., {Collet}, R., {Amarsi}, A.~M., {et~al.} 2017, \apj, 847, 15, \dodoi{10.3847/1538-4357/aa88cb}

\bibitem[{{Bergemann} \& {Gehren}(2008)}]{Bergemann2008Mn}
{Bergemann}, M., \& {Gehren}, T. 2008, \aap, 492, 823, \dodoi{10.1051/0004-6361:200810098}

\bibitem[{{Bergemann} {et~al.}(2012){Bergemann}, {Lind}, {Collet}, {Magic}, \& {Asplund}}]{bergemann2012}
{Bergemann}, M., {Lind}, K., {Collet}, R., {Magic}, Z., \& {Asplund}, M. 2012, \mnras, 427, 27, \dodoi{10.1111/j.1365-2966.2012.21687.x}

\bibitem[{{Bisterzo} {et~al.}(2010){Bisterzo}, {Gallino}, {Straniero}, {Cristallo}, \& {K{\"a}ppeler}}]{bisterzo}
{Bisterzo}, S., {Gallino}, R., {Straniero}, O., {Cristallo}, S., \& {K{\"a}ppeler}, F. 2010, \mnras, 404, 1529, \dodoi{10.1111/j.1365-2966.2010.16369.x}

\bibitem[{{Bragaglia} {et~al.}(2010){Bragaglia}, {Carretta}, {Gratton}, {Lucatello}, {Milone}, {Piotto}, {D'Orazi}, {Cassisi}, {Sneden}, \& {Bedin}}]{brag2010}
{Bragaglia}, A., {Carretta}, E., {Gratton}, R.~G., {et~al.} 2010, \apjl, 720, L41, \dodoi{10.1088/2041-8205/720/1/L41}

\bibitem[{Brown \& Forsythe(1974)}]{brownfs}
Brown, M.~B., \& Forsythe, A.~B. 1974, Journal of the American Statistical Association, 69, 364.
\newblock \url{http://www.jstor.org/stable/2285659}

\bibitem[{{Carballo-Bello} {et~al.}(2018){Carballo-Bello}, {Mart{\'\i}nez-Delgado}, {Navarrete}, {Catelan}, {Mu{\~n}oz}, {Antoja}, \& {Sollima}}]{julio2018}
{Carballo-Bello}, J.~A., {Mart{\'\i}nez-Delgado}, D., {Navarrete}, C., {et~al.} 2018, \mnras, 474, 683, \dodoi{10.1093/mnras/stx2767}

\bibitem[{Carretta(2015)}]{Carretta-ngc2808-2015}
Carretta, E. 2015, The Astrophysical Journal, 810, 148, \dodoi{10.1088/0004-637X/810/2/148}

\bibitem[{{Carretta}(2019)}]{car2019}
{Carretta}, E. 2019, \aap, 624, A24, \dodoi{10.1051/0004-6361/201935110}

\bibitem[{{Carretta} {et~al.}(2009{\natexlab{a}}){Carretta}, {Bragaglia}, {Gratton}, \& {Lucatello}}]{carretta2009}
{Carretta}, E., {Bragaglia}, A., {Gratton}, R., \& {Lucatello}, S. 2009{\natexlab{a}}, \aap, 505, 139, \dodoi{10.1051/0004-6361/200912097}

\bibitem[{{Carretta} {et~al.}(2010){Carretta}, {Bragaglia}, {Gratton}, {Recio-Blanco}, {Lucatello}, {D'Orazi}, \& {Cassisi}}]{carretta2010x}
{Carretta}, E., {Bragaglia}, A., {Gratton}, R.~G., {et~al.} 2010, \aap, 516, A55, \dodoi{10.1051/0004-6361/200913451}

\bibitem[{{Carretta} {et~al.}(2017){Carretta}, {Bragaglia}, {Lucatello}, {D'Orazi}, {Gratton}, {Donati}, {Sollima}, \& {Sneden}}]{carretta2017}
{Carretta}, E., {Bragaglia}, A., {Lucatello}, S., {et~al.} 2017, \aap, 600, A118, \dodoi{10.1051/0004-6361/201630004}

\bibitem[{{Carretta} {et~al.}(2009{\natexlab{b}}){Carretta}, {Bragaglia}, {Gratton}, {Lucatello}, {Catanzaro}, {Leone}, {Bellazzini}, {Claudi}, {D'Orazi}, {Momany}, {Ortolani}, {Pancino}, {Piotto}, {Recio-Blanco}, \& {Sabbi}}]{carretta2009b}
{Carretta}, E., {Bragaglia}, A., {Gratton}, R.~G., {et~al.} 2009{\natexlab{b}}, \aap, 505, 117, \dodoi{10.1051/0004-6361/200912096}

\bibitem[{{Castelli} \& {Kurucz}(2004)}]{castelli2004}
{Castelli}, F., \& {Kurucz}, R.~L. 2004

\bibitem[{{Cayrel}(1988)}]{Cayrel88}
{Cayrel}, R. 1988, in The Impact of Very High S/N Spectroscopy on Stellar Physics, ed. G.~{Cayrel de Strobel} \& M.~{Spite}, Vol. 132, 345

\bibitem[{{Cummings} {et~al.}(2017){Cummings}, {Geisler}, \& {Villanova}}]{cummings2017}
{Cummings}, J.~D., {Geisler}, D., \& {Villanova}, S. 2017, \aj, 153, 192, \dodoi{10.3847/1538-3881/aa63e5}

\bibitem[{D'Antona {et~al.}(2019)D'Antona, Ventura, Marino, Milone, Tailo, Criscienzo, \& Vesperini}]{dantona2019}
D'Antona, F., Ventura, P., Marino, A.~F., {et~al.} 2019, The Astrophysical Journal, 871, L19, \dodoi{10.3847/2041-8213/aafbec}

\bibitem[{{De Lucia} {et~al.}(2024){De Lucia}, {Kruijssen}, {Trujillo-Gomez}, {Hirschmann}, \& {Xie}}]{lucia2024}
{De Lucia}, G., {Kruijssen}, J.~M.~D., {Trujillo-Gomez}, S., {Hirschmann}, M., \& {Xie}, L. 2024, \mnras, 530, 2760, \dodoi{10.1093/mnras/stae1006}

\bibitem[{{Denissenkov} {et~al.}(1998){Denissenkov}, {Da Costa}, {Norris}, \& {Weiss}}]{den97}
{Denissenkov}, P.~A., {Da Costa}, G.~S., {Norris}, J.~E., \& {Weiss}, A. 1998, \aap, 333, 926, \dodoi{10.48550/arXiv.astro-ph/9707266}

\bibitem[{{Dondoglio} {et~al.}(2022){Dondoglio}, {Milone}, {Renzini}, {Vesperini}, {Lagioia}, {Marino}, {Bellini}, {Carlos}, {Cordoni}, {Jang}, {Legnardi}, {Libralato}, {Mohandasan}, {D'Antona}, {Martorano}, {Muratore}, \& {Tailo}}]{don2022}
{Dondoglio}, E., {Milone}, A.~P., {Renzini}, A., {et~al.} 2022, \apj, 927, 207, \dodoi{10.3847/1538-4357/ac5046}

\bibitem[{{Dotter} {et~al.}(2015){Dotter}, {Ferguson}, {Conroy}, {Milone}, {Marino}, \& {Yong}}]{dotter2014}
{Dotter}, A., {Ferguson}, J.~W., {Conroy}, C., {et~al.} 2015, \mnras, 446, 1641, \dodoi{10.1093/mnras/stu2170}

\bibitem[{{Elmegreen}(2018)}]{elme2018}
{Elmegreen}, B.~G. 2018, \apj, 869, 119, \dodoi{10.3847/1538-4357/aaed45}

\bibitem[{Ernandes {et~al.}(2023)Ernandes, Castro, Barbuy, Spite, Hill, Castilho, \& Evans}]{Ernandes2023}
Ernandes, H., Castro, M.~J., Barbuy, B., {et~al.} 2023, Monthly Notices of the Royal Astronomical Society, 524, 656, \dodoi{10.1093/mnras/stad1764}

\bibitem[{{Ezzeddine} {et~al.}(2016){Ezzeddine}, {Merle}, \& {Plez}}]{Ezzeddine2016a}
{Ezzeddine}, R., {Merle}, T., \& {Plez}, B. 2016, Astronomische Nachrichten, 337, 850, \dodoi{10.1002/asna.201612384}

\bibitem[{Ezzeddine {et~al.}(2020)Ezzeddine, Rasmussen, Frebel, Chiti, Hinojisa, Placco, Ji, Beers, Hansen, Roederer, Sakari, \& Melendez}]{rpa3}
Ezzeddine, R., Rasmussen, K., Frebel, A., {et~al.} 2020, The Astrophysical Journal, 898, 150, \dodoi{10.3847/1538-4357/ab9d1a}

\bibitem[{{Forbes} {et~al.}(2018){Forbes}, {Bastian}, {Gieles}, {Crain}, {Kruijssen}, {Larsen}, {Ploeckinger}, {Agertz}, {Trenti}, {Ferguson}, {Pfeffer}, \& {Gnedin}}]{forbes2018}
{Forbes}, D.~A., {Bastian}, N., {Gieles}, M., {et~al.} 2018, Proceedings of the Royal Society of London Series A, 474, 20170616, \dodoi{10.1098/rspa.2017.0616}

\bibitem[{{Frebel} {et~al.}(2013){Frebel}, {Casey}, {Jacobson}, \& {Yu}}]{frebel2013}
{Frebel}, A., {Casey}, A.~R., {Jacobson}, H.~R., \& {Yu}, Q. 2013, \apj, 769, 57, \dodoi{10.1088/0004-637X/769/1/57}

\bibitem[{{Gratton} {et~al.}(2019){Gratton}, {Bragaglia}, {Carretta}, {D'Orazi}, {Lucatello}, \& {Sollima}}]{gratton2019}
{Gratton}, R., {Bragaglia}, A., {Carretta}, E., {et~al.} 2019, \aapr, 27, 8, \dodoi{10.1007/s00159-019-0119-3}

\bibitem[{{Gratton} {et~al.}(2004){Gratton}, {Sneden}, \& {Carretta}}]{gratton2004}
{Gratton}, R., {Sneden}, C., \& {Carretta}, E. 2004, \araa, 42, 385, \dodoi{10.1146/annurev.astro.42.053102.133945}

\bibitem[{{Gratton} {et~al.}(2012){Gratton}, {Carretta}, \& {Bragaglia}}]{gratton2012}
{Gratton}, R.~G., {Carretta}, E., \& {Bragaglia}, A. 2012, \aapr, 20, 50, \dodoi{10.1007/s00159-012-0050-3}

\bibitem[{{Gratton} {et~al.}(2011){Gratton}, {Lucatello}, {Carretta}, {Bragaglia}, {D'Orazi}, \& {Momany}}]{gratton2011}
{Gratton}, R.~G., {Lucatello}, S., {Carretta}, E., {et~al.} 2011, \aap, 534, A123, \dodoi{10.1051/0004-6361/201117690}

\bibitem[{{Gratton} {et~al.}(2001){Gratton}, {Bonifacio}, {Bragaglia}, {Carretta}, {Castellani}, {Centurion}, {Chieffi}, {Claudi}, {Clementini}, {D'Antona}, {Desidera}, {Fran{\c c}ois}, {Grundahl}, {Lucatello}, {Molaro}, {Pasquini}, {Sneden}, {Spite}, \& {Straniero}}]{gratton2001}
{Gratton}, R.~G., {Bonifacio}, P., {Bragaglia}, A., {et~al.} 2001, \aap, 369, 87, \dodoi{10.1051/0004-6361:20010144}

\bibitem[{{Holmbeck} {et~al.}(2018){Holmbeck}, {Beers}, {Roederer}, {Placco}, {Hansen}, {Sakari}, {Sneden}, {Liu}, {Lee}, {Cowan}, \& {Frebel}}]{holmbeck2018}
{Holmbeck}, E.~M., {Beers}, T.~C., {Roederer}, I.~U., {et~al.} 2018, \apjl, 859, L24, \dodoi{10.3847/2041-8213/aac722}

\bibitem[{{Ishchenko} {et~al.}(2023){Ishchenko}, {Sobolenko}, {Berczik}, {Khoperskov}, {Omarov}, {Sobodar}, \& {Makukov}}]{is2023}
{Ishchenko}, M., {Sobolenko}, M., {Berczik}, P., {et~al.} 2023, \aap, 673, A152, \dodoi{10.1051/0004-6361/202245117}

\bibitem[{{James} {et~al.}(2004){James}, {Fran{\c{c}}ois}, {Bonifacio}, {Carretta}, {Gratton}, \& {Spite}}]{james2004}
{James}, G., {Fran{\c{c}}ois}, P., {Bonifacio}, P., {et~al.} 2004, \aap, 427, 825, \dodoi{10.1051/0004-6361:20041512}

\bibitem[{{Janes} \& {Heasley}(1988)}]{janes88}
{Janes}, K.~A., \& {Heasley}, J.~N. 1988, \aj, 95, 762, \dodoi{10.1086/114675}

\bibitem[{{Ji} {et~al.}(2016){Ji}, {Frebel}, {Simon}, \& {Chiti}}]{ji2016b}
{Ji}, A.~P., {Frebel}, A., {Simon}, J.~D., \& {Chiti}, A. 2016, The Astrophysical Journal, 830, 93, \dodoi{10.3847/0004-637X/830/2/93}

\bibitem[{{Johnson} {et~al.}(2015){Johnson}, {Rich}, {Pilachowski}, {Caldwell}, {Mateo}, {Bailey}, \& {Crane}}]{Johnson2015}
{Johnson}, C.~I., {Rich}, R.~M., {Pilachowski}, C.~A., {et~al.} 2015, \aj, 150, 63, \dodoi{10.1088/0004-6256/150/2/63}

\bibitem[{{Kalari} {et~al.}(2024){Kalari}, {Diaz}, {Robertson}, {McConnachie}, {Ireland}, {Salinas}, {Young}, {Simpson}, {Hayes}, {Nielsen}, {Burley}, {Pazder}, {Gomez-Jimenez}, {Martioli}, {Howell}, {Jeong}, {Juneau}, {Ruiz-Carmona}, {Margheim}, {Sheinis}, {Anthony}, {Baker}, {Berg}, {Cao}, {Chapin}, {Chin}, {Chiboucas}, {Churilov}, {Deibert}, {Densmore}, {Dunn}, {Edgar}, {Heo}, {Henderson}, {Farrell}, {Font}, {Firpo}, {Fuentes}, {Labrie}, {Lambert}, {Lawrence}, {Lothrop}, {McDermid}, {Miller}, {Perez}, {Placco}, {Prado}, {Quiroz}, {Ramos}, {Rutten}, {Silva}, {Thomas-Osip}, {Urrutia}, {Vacca}, {Venn}, {Waller}, {Waller}, {White}, {Xu}, \& {Zhelem}}]{Kalari2024}
{Kalari}, V.~M., {Diaz}, R.~J., {Robertson}, G., {et~al.} 2024, \aj, 168, 208, \dodoi{10.3847/1538-3881/ad72ed}

\bibitem[{{Kim} \& {Lee}(2018)}]{kiml18}
{Kim}, J.~J., \& {Lee}, Y.-W. 2018, \apj, 869, 35, \dodoi{10.3847/1538-4357/aaec67}

\bibitem[{{Kimm} {et~al.}(2016){Kimm}, {Cen}, {Rosdahl}, \& {Yi}}]{kimm2016}
{Kimm}, T., {Cen}, R., {Rosdahl}, J., \& {Yi}, S.~K. 2016, \apj, 823, 52, \dodoi{10.3847/0004-637X/823/1/52}

\bibitem[{{Kirby} {et~al.}(2023){Kirby}, {Ji}, \& {Kovalev}}]{kirby23}
{Kirby}, E.~N., {Ji}, A.~P., \& {Kovalev}, M. 2023, \apj, 958, 45, \dodoi{10.3847/1538-4357/acf309}

\bibitem[{Kobayashi {et~al.}(2006)Kobayashi, Umeda, Nomoto, Tominaga, \& Ohkubo}]{kobayashi2006}
Kobayashi, C., Umeda, H., Nomoto, K., Tominaga, N., \& Ohkubo, T. 2006, The Astrophysical Journal, 653, 1145

\bibitem[{{Korotin} {et~al.}(2015){Korotin}, {Andrievsky}, {Hansen}, {Caffau}, {Bonifacio}, {Spite}, {Spite}, \& {Fran{\c{c}}ois}}]{korotin2015nlte}
{Korotin}, S.~A., {Andrievsky}, S.~M., {Hansen}, C.~J., {et~al.} 2015, \aap, 581, A70, \dodoi{10.1051/0004-6361/201526558}

\bibitem[{{Kraft}(1979)}]{kraftb}
{Kraft}, R.~P. 1979, \araa, 17, 309, \dodoi{10.1146/annurev.aa.17.090179.001521}

\bibitem[{{Kraft} {et~al.}(1979){Kraft}, {Trefzger}, \& {Suntzeff}}]{krafta}
{Kraft}, R.~P., {Trefzger}, C.~F., \& {Suntzeff}, N. 1979, in IAU Symposium, Vol.~84, The Large-Scale Characteristics of the Galaxy, ed. W.~B. {Burton}, 463--473

\bibitem[{{Labrie} {et~al.}(2023){Labrie}, {Simpson}, {Cardenes}, {Turner}, {Soraisam}, {Quint}, {Oberdorf}, {Placco}, {Berke}, {Smirnova}, {Conseil}, {Vacca}, \& {Thomas-Osip}}]{dragons}
{Labrie}, K., {Simpson}, C., {Cardenes}, R., {et~al.} 2023, Research Notes of the American Astronomical Society, 7, 214, \dodoi{10.3847/2515-5172/ad0044}

\bibitem[{Leaman(2012)}]{Leaman_2012}
Leaman, R. 2012, The Astronomical Journal, 144, 183, \dodoi{10.1088/0004-6256/144/6/183}

\bibitem[{Lee(2010)}]{Lee2010}
Lee, J.-W. 2010, Monthly Notices of the Royal Astronomical Society: Letters, 405, L36, \dodoi{10.1111/j.1745-3933.2010.00854.x}

\bibitem[{{Lind} {et~al.}(2011){Lind}, {Asplund}, {Barklem}, \& {Belyaev}}]{lind2011}
{Lind}, K., {Asplund}, M., {Barklem}, P.~S., \& {Belyaev}, A.~K. 2011, \aap, 528, A103, \dodoi{10.1051/0004-6361/201016095}

\bibitem[{{Lind} {et~al.}(2012){Lind}, {Bergemann}, \& {Asplund}}]{lind2012}
{Lind}, K., {Bergemann}, M., \& {Asplund}, M. 2012, \mnras, 427, 50, \dodoi{10.1111/j.1365-2966.2012.21686.x}

\bibitem[{{Marigo} {et~al.}(2017){Marigo}, {Girardi}, {Bressan}, {Rosenfield}, {Aringer}, {Chen}, {Dussin}, {Nanni}, {Pastorelli}, {Rodrigues}, {Trabucchi}, {Bladh}, {Dalcanton}, {Groenewegen}, {Montalb{\'a}n}, \& {Wood}}]{Marigo2017}
{Marigo}, P., {Girardi}, L., {Bressan}, A., {et~al.} 2017, \apj, 835, 77, \dodoi{10.3847/1538-4357/835/1/77}

\bibitem[{{Martin} {et~al.}(2004){Martin}, {Ibata}, {Bellazzini}, {Irwin}, {Lewis}, \& {Dehnen}}]{martin2004}
{Martin}, N.~F., {Ibata}, R.~A., {Bellazzini}, M., {et~al.} 2004, \mnras, 348, 12, \dodoi{10.1111/j.1365-2966.2004.07331.x}

\bibitem[{{Mashonkina} {et~al.}(2023){Mashonkina}, {Pakhomov}, {Sitnova}, {Smogorzhevskii}, {Jablonka}, \& {Hill}}]{Mashonkina2023}
{Mashonkina}, L., {Pakhomov}, Y., {Sitnova}, T., {et~al.} 2023, \mnras, 524, 3526, \dodoi{10.1093/mnras/stad2114}

\bibitem[{{Massari} {et~al.}(2019){Massari}, {Koppelman}, \& {Helmi}}]{massari19}
{Massari}, D., {Koppelman}, H.~H., \& {Helmi}, A. 2019, \aap, 630, L4, \dodoi{10.1051/0004-6361/201936135}

\bibitem[{McWilliam {et~al.}(1992)McWilliam, Geisler, \& Rich}]{mc92}
McWilliam, A., Geisler, D., \& Rich, R.~M. 1992, Publications of the Astronomical Society of the Pacific, 104, 1193.
\newblock \url{http://www.jstor.org/stable/40679982}

\bibitem[{{M{\'e}sz{\'a}ros} {et~al.}(2020){M{\'e}sz{\'a}ros}, {Masseron}, {Garc{\'\i}a-Hern{\'a}ndez}, {Allende Prieto}, {Beers}, {Bizyaev}, {Chojnowski}, {Cohen}, {Cunha}, {Dell'Agli}, {Ebelke}, {Fern{\'a}ndez-Trincado}, {Frinchaboy}, {Geisler}, {Hasselquist}, {Hearty}, {Holtzman}, {Johnson}, {Lane}, {Lacerna}, {Longa-Pe{\~n}a}, {Majewski}, {Martell}, {Minniti}, {Nataf}, {Nidever}, {Pan}, {Schiavon}, {Shetrone}, {Smith}, {Sobeck}, {Stringfellow}, {Szigeti}, {Tang}, {Wilson}, \& {Zamora}}]{meszaros2020}
{M{\'e}sz{\'a}ros}, S., {Masseron}, T., {Garc{\'\i}a-Hern{\'a}ndez}, D.~A., {et~al.} 2020, \mnras, 492, 1641, \dodoi{10.1093/mnras/stz3496}

\bibitem[{{Milone} \& {Marino}(2022)}]{milone2022}
{Milone}, A.~P., \& {Marino}, A.~F. 2022, Universe, 8, 359, \dodoi{10.3390/universe8070359}

\bibitem[{{Milone} {et~al.}(2012){Milone}, {Piotto}, {Bedin}, {Aparicio}, {Anderson}, {Sarajedini}, {Marino}, {Moretti}, {Davies}, {Chaboyer}, {Dotter}, {Hempel}, {Mar{\'\i}n-Franch}, {Majewski}, {Paust}, {Reid}, {Rosenberg}, \& {Siegel}}]{milone2012}
{Milone}, A.~P., {Piotto}, G., {Bedin}, L.~R., {et~al.} 2012, \aap, 540, A16, \dodoi{10.1051/0004-6361/201016384}

\bibitem[{Moreno {et~al.}(2014)Moreno, Pichardo, \& Velázquez}]{Moreno2014}
Moreno, E., Pichardo, B., \& Velázquez, H. 2014, The Astrophysical Journal, 793, 110, \dodoi{10.1088/0004-637X/793/2/110}

\bibitem[{{Mu{\~n}oz} {et~al.}(2017){Mu{\~n}oz}, {Villanova}, {Geisler}, {Saviane}, {Dias}, {Cohen}, \& {Mauro}}]{munoz2017}
{Mu{\~n}oz}, C., {Villanova}, S., {Geisler}, D., {et~al.} 2017, \aap, 605, A12, \dodoi{10.1051/0004-6361/201730468}

\bibitem[{{Mucciarelli} {et~al.}(2021){Mucciarelli}, {Bellazzini}, \& {Massari}}]{muc2021gaia}
{Mucciarelli}, A., {Bellazzini}, M., \& {Massari}, D. 2021, \aap, 653, A90, \dodoi{10.1051/0004-6361/202140979}

\bibitem[{{Mucciarelli} \& {Bonifacio}(2020)}]{Mucboni2021}
{Mucciarelli}, A., \& {Bonifacio}, P. 2020, \aap, 640, A87, \dodoi{10.1051/0004-6361/202037703}

\bibitem[{{Mucciarelli} {et~al.}(2018){Mucciarelli}, {Lapenna}, {Ferraro}, \& {Lanzoni}}]{mucci2018}
{Mucciarelli}, A., {Lapenna}, E., {Ferraro}, F.~R., \& {Lanzoni}, B. 2018, \apj, 859, 75, \dodoi{10.3847/1538-4357/aaba80}

\bibitem[{{Nordlander} \& {Lind}(2017)}]{nordlander2017}
{Nordlander}, T., \& {Lind}, K. 2017, \aap, 607, A75, \dodoi{10.1051/0004-6361/201730427}

\bibitem[{{Pancino} {et~al.}(2017){Pancino}, {Romano}, {Tang}, {Tautvai{\v{s}}ien{\.{e}}}, {Casey}, {Gruyters}, {Geisler}, {San Roman}, {Randich}, {Alfaro}, {Bragaglia}, {Flaccomio}, {Korn}, {Recio-Blanco}, {Smiljanic}, {Carraro}, {Bayo}, {Costado}, {Damiani}, {Jofr{\'e}}, {Lardo}, {de Laverny}, {Monaco}, {Morbidelli}, {Sbordone}, {Sousa}, \& {Villanova}}]{pan2017}
{Pancino}, E., {Romano}, D., {Tang}, B., {et~al.} 2017, \aap, 601, A112, \dodoi{10.1051/0004-6361/201730474}

\bibitem[{{Pasquato} {et~al.}(2009){Pasquato}, {Trenti}, {De Marchi}, {Gill}, {Hamilton}, {Miller}, {Stiavelli}, \& {van der Marel}}]{pas2009}
{Pasquato}, M., {Trenti}, M., {De Marchi}, G., {et~al.} 2009, \apj, 699, 1511, \dodoi{10.1088/0004-637X/699/2/1511}

\bibitem[{{Piotto} {et~al.}(2007){Piotto}, {Bedin}, {Anderson}, {King}, {Cassisi}, {Milone}, {Villanova}, {Pietrinferni}, \& {Renzini}}]{piotto2007}
{Piotto}, G., {Bedin}, L.~R., {Anderson}, J., {et~al.} 2007, \apjl, 661, L53, \dodoi{10.1086/518503}

\bibitem[{{Placco} {et~al.}(2014){Placco}, {Frebel}, {Beers}, \& {Stancliffe}}]{placco2014}
{Placco}, V.~M., {Frebel}, A., {Beers}, T.~C., \& {Stancliffe}, R.~J. 2014, \apj, 797, 21, \dodoi{10.1088/0004-637X/797/1/21}

\bibitem[{{Placco} {et~al.}(2021){Placco}, {Sneden}, {Roederer}, {Lawler}, {Den Hartog}, {Hejazi}, {Maas}, \& {Bernath}}]{linemake_vini}
{Placco}, V.~M., {Sneden}, C., {Roederer}, I.~U., {et~al.} 2021, Research Notes of the American Astronomical Society, 5, 92, \dodoi{10.3847/2515-5172/abf651}

\bibitem[{{Placco} {et~al.}(2024){Placco}, {Herrera}, {Merino}, {US National Gemini Office}, {Hirst}, {Labrie}, {Simpson}, {Turner}, {Vacca}, {Gemini Science User Support Department}, {Deibert}, {Diaz}, {Heo}, {Kalari}, {Reggiani}, {Rodriguez}, {Ruiz-Carmona}, {Thomas-Osip}, \& {GHOST Instrument Team}}]{Placco2024RNAAS}
{Placco}, V.~M., {Herrera}, D., {Merino}, B.~M., {et~al.} 2024, Research Notes of the American Astronomical Society, 8, 312, \dodoi{10.3847/2515-5172/ad9f5f}

\bibitem[{{Rantakyro} {et~al.}(2024){Rantakyro}, {Kalari}, \& {Placco}}]{ghost}
{Rantakyro}, F., {Kalari}, V., \& {Placco}, V. 2024, The NOIRLab Mirror, 6, 28

\bibitem[{Roederer(2011)}]{Roederer2011}
Roederer, I.~U. 2011, The Astrophysical Journal Letters, 732, L17, \dodoi{10.1088/2041-8205/732/1/L17}

\bibitem[{{Roederer} {et~al.}(2018){Roederer}, {Hattori}, \& {Valluri}}]{roederer2018}
{Roederer}, I.~U., {Hattori}, K., \& {Valluri}, M. 2018, \aj, 156, 179, \dodoi{10.3847/1538-3881/aadd9c}

\bibitem[{{Roederer} \& {Thompson}(2015)}]{Roe2015}
{Roederer}, I.~U., \& {Thompson}, I.~B. 2015, \mnras, 449, 3889, \dodoi{10.1093/mnras/stv546}

\bibitem[{{Sills} \& {Glebbeek}(2010)}]{sillis2010}
{Sills}, A., \& {Glebbeek}, E. 2010, \mnras, 407, 277, \dodoi{10.1111/j.1365-2966.2010.16876.x}

\bibitem[{{Simpson} {et~al.}(2017){Simpson}, {Martell}, \& {Navin}}]{simpson2017}
{Simpson}, J.~D., {Martell}, S.~L., \& {Navin}, C.~A. 2017, \mnras, 465, 1123, \dodoi{10.1093/mnras/stw2781}

\bibitem[{{Sneden} {et~al.}(2008){Sneden}, {Cowan}, \& {Gallino}}]{sneden2008}
{Sneden}, C., {Cowan}, J.~J., \& {Gallino}, R. 2008, \araa, 46, 241, \dodoi{10.1146/annurev.astro.46.060407.145207}

\bibitem[{Sneden {et~al.}(2003)Sneden, Cowan, Lawler, Ivans, Burles, Beers, Primas, Hill, Truran, Fuller, Pfeiffer, \& Kratz}]{Sneden2003}
Sneden, C., Cowan, J.~J., Lawler, J.~E., {et~al.} 2003, The Astrophysical Journal, 591, 936, \dodoi{10.1086/375491}

\bibitem[{{Sneden}(1973)}]{sneden1973}
{Sneden}, C.~A. 1973, PhD thesis, University of Texas at Austin

\bibitem[{{Sobeck} {et~al.}(2011){Sobeck}, {Kraft}, {Sneden}, {Preston}, {Cowan}, {Smith}, {Thompson}, {Shectman}, \& {Burley}}]{sobeck2011}
{Sobeck}, J.~S., {Kraft}, R.~P., {Sneden}, C., {et~al.} 2011, \aj, 141, 175, \dodoi{10.1088/0004-6256/141/6/175}

\bibitem[{{Truran} {et~al.}(2002){Truran}, {Cowan}, {Pilachowski}, \& {Sneden}}]{truran2002}
{Truran}, J.~W., {Cowan}, J.~J., {Pilachowski}, C.~A., \& {Sneden}, C. 2002, \pasp, 114, 1293, \dodoi{10.1086/344585}

\bibitem[{{Tsujimoto} \& {Shigeyama}(2014)}]{tsuji1}
{Tsujimoto}, T., \& {Shigeyama}, T. 2014, \apjl, 795, L18, \dodoi{10.1088/2041-8205/795/1/L18}

\bibitem[{{Vasiliev} \& {Baumgardt}(2021)}]{vasiliev2021}
{Vasiliev}, E., \& {Baumgardt}, H. 2021, \mnras, 505, 5978, \dodoi{10.1093/mnras/stab1475}

\bibitem[{Venn {et~al.}(2004)Venn, Irwin, Shetrone, Tout, Hill, \& Tolstoy}]{venn2004}
Venn, K.~A., Irwin, M., Shetrone, M.~D., {et~al.} 2004, The Astronomical Journal, 128, 1177

\bibitem[{{Walker} {et~al.}(2006){Walker}, {Mateo}, {Olszewski}, {Bernstein}, {Wang}, \& {Woodroofe}}]{Walker2006}
{Walker}, M.~G., {Mateo}, M., {Olszewski}, E.~W., {et~al.} 2006, \aj, 131, 2114, \dodoi{10.1086/500193}

\bibitem[{Wilcox(2017)}]{wilcox2017}
Wilcox, R.~R. 2017, Introduction to Robust Estimation and Hypothesis Testing, 3rd edn. (Academic Press)

\bibitem[{{Zevin} {et~al.}(2019){Zevin}, {Kremer}, {Siegel}, {Coughlin}, {Tsang}, {Berry}, \& {Kalogera}}]{Zevin19}
{Zevin}, M., {Kremer}, K., {Siegel}, D.~M., {et~al.} 2019, \apj, 886, 4, \dodoi{10.3847/1538-4357/ab498b}

\end{thebibliography}
\bibliographystyle{aasjournal}
\begin{longrotatetable}
\begin{flushleft}
\begin{deluxetable}{l r r r r r r r r r r r r r r r r}
\tablewidth{0pt}
\tabletypesize{\tiny}
\tablecaption{\label{tab:abund_fe_peak}Complete Elemental Abundances [X/Fe]}
\tablehead{
\colhead{Element} & \colhead{11} & \colhead{12} & \colhead{13} & \colhead{14} & \colhead{581} & \colhead{582} & \colhead{591} & \colhead{592} & \colhead{601} & \colhead{602} & \colhead{611} & \colhead{612} & \colhead{981}
}
\startdata
C &$-0.7\pm 0.09$ &$-0.83\pm 0.11$ &$-1.24\pm 0.08$ &$-1.17\pm 0.07$ &$-1.02\pm 0.09$ &$-1.27\pm 0.10$ &$-0.98\pm 0.12$ &$-1.1\pm 0.09$ &$-0.67\pm 0.13$ &$-0.98\pm 0.10$ &$-0.89\pm 0.11$ &$-0.94\pm 0.09$ &$-0.93\pm 0.12$ \\
N &$0.66\pm 0.14$ &$0.51\pm 0.16$ &$1.34\pm 0.13$ &$1.20\pm 0.11$ &$1.10\pm 0.14$ &$1.05\pm 0.13$ &$0.55\pm 0.12$ &$-0.010.47\pm 0.12$ &$0.65\pm 0.13$ &$0.95\pm 0.11$ &$0.80\pm 0.14$ &$0.50\pm 0.13$ &$0.61\pm 0.12$ \\
Na &$0.23\pm 0.13$ &$0.07\pm 0.11$ &$0.59\pm 0.10$ &$0.54\pm 0.08$ &$0.54\pm 0.10$ &$0.54\pm 0.11$ &$0\pm 0.12$ &$-0.11\pm 0.10$ &$0.21\pm 0.14$ &$0.37\pm 0.13$ &$0.36\pm 0.12$ &$-0.07\pm 0.10$ &$0.02\pm 0.13$ \\
Mg\ &$0.37\pm 0.09$ &$0.57\pm 0.08$ &$0.21\pm 0.11$ &$0.15\pm 0.10$ &$0.19\pm 0.12$ &$0.13\pm 0.09$ &$0.37\pm 0.10$ &$0.55\pm 0.11$ &$0.32\pm 0.08$ &$0.23\pm 0.12$ &$0.33\pm 0.10$ &$0.32\pm 0.13$ &$0.36\pm 0.14$ \\
Al &$-0.03\pm 0.15$ &$-0.64\pm 0.17$ &$0.27\pm 0.18$ &$0.22\pm 0.14$ &$0.16\pm 0.18$ &$0.02\pm 0.17$ &$-0.53\pm 0.14$ &$-0.66\pm 0.15$ &$-0.57\pm 0.18$ &$-0.09\pm 0.12$ &$-0.28\pm 0.19$ &$-0.71\pm 0.16$ &$-0.58\pm 0.13$ \\
Si &$0.2\pm 0.14$ &$0.13\pm 0.15$ &$0.05\pm 0.11$ &$0.07\pm 0.15$ &$0.23\pm 0.12$ &$0.56\pm 0.17$ &$0.04\pm 0.15$ &$999\pm 0.13$ &$0.21\pm 0.14$ &$0.07\pm 0.16$ &$0.23\pm 0.12$ &$0.38\pm 0.18$ &$0.19\pm 0.16$ \\
Ca &$0.4\pm 0.13$ &$0.45\pm 0.11$ &$0.38\pm 0.14$ &$0.39\pm 0.11$ &$0.42\pm 0.12$ &$0.41\pm 0.13$ &$0.43\pm 0.10$ &$0.45\pm 0.14$ &$0.44\pm 0.11$ &$0.44\pm 0.10$ &$0.4\pm 0.10$ &$0.44\pm 0.13$ &$0.44\pm 0.15$ \\
Sc &$0.05\pm 0.10$ &$0.11\pm 0.09$ &$0.24\pm 0.08$ &$0.18\pm 0.11$ &$0.19\pm 0.12$ &$0.28\pm 0.10$ &$0.02\pm 0.11$ &$0.46\pm 0.13$ &$-0.02\pm 0.12$ &$0.18\pm 0.08$ &$0.05\pm 0.11$ &$0.19\pm 0.13$ &$0.16\pm 0.10$ \\
Ti &$0.25\pm 0.11$ &$0.32\pm 0.13$ &$0.22\pm 0.13$ &$0.26\pm 0.12$ &$0.36\pm 0.16$ &$0.4\pm 0.12$ &$0.24\pm 0.11$ &$0.45\pm 0.12$ &$0.25\pm 0.14$ &$0.4\pm 0.17$ &$0.25\pm 0.14$ &$0.38\pm 0.13$ &$0.31\pm 0.11$ \\
V &$0.04\pm 0.17$ &$0.05\pm 0.14$ &$0.05\pm 0.16$ &$0.14\pm 0.14$ &$0.25\pm 0.13$ &$0.39\pm 0.15$ &$0.07\pm 0.16$ &$0.24\pm 0.14$ &$0.15\pm 0.18$ &$0.11\pm 0.16$ &$0.05\pm 0.15$ &$0.27\pm 0.14$ &$0.19\pm 0.16$ \\
Cr &$0.01\pm 0.15$ &$0.04\pm 0.14$ &$0.03\pm 0.11$ &$0.05\pm 0.13$ &$0.02\pm 0.15$ &$0.16\pm 0.12$ &$0\pm 0.11$ &$0.08\pm 0.14$ &$0.01\pm 0.16$ &$0.11\pm 0.12$ &$0\pm 0.11$ &$0.15\pm 0.10$ &$0.02\pm 0.13$ \\
Mn &$-0.18\pm 0.14$ &$-0.14\pm 0.12$ &$-0.19\pm 0.11$ &$-0.14\pm 0.13$ &$-0.12\pm 0.15$ &$0\pm 0.14$ &$-0.01\pm 0.16$ &$-0.04\pm 0.12$ &$-0.16\pm 0.14$ &$-0.18\pm 0.13$ &$-0.16\pm 0.14$ &$-0.24\pm 0.15$ &$-0.08\pm 0.12$ \\
Co &$-0.14\pm 0.11$ &$-0.22\pm 0.13$ &$-0.29\pm 0.13$ &$0.17\pm 0.16$ &$-0.05\pm 0.14$ &$0.34\pm 0.15$ &$-0.13\pm 0.13$ &$-0.1\pm 0.12$ &$-0.15\pm 0.14$ &$-0.27\pm 0.15$ &$-0.33\pm 0.16$ &$0.1\pm 0.18$ &$-0.02\pm 0.13$ \\
Ni &$0.03\pm 0.14$ &$-0.03\pm 0.12$ &$-0.03\pm 0.14$ &$-0.06\pm 0.13$ &$-0.01\pm 0.12$ &$-0.07\pm 0.13$ &$-0.11\pm 0.12$ &$-0.1\pm 0.14$ &$0.04\pm 0.15$ &$-0.05\pm 0.14$ &$-0.03\pm 0.13$ &$-0.05\pm 0.11$ &$-0.1\pm 0.16$ \\
Cu &$-0.56\pm 0.12$ &$-0.39\pm 0.13$ &$-0.48\pm 0.13$ &$-0.49\pm 0.12$ &$-0.44\pm 0.15$ &$-0.53\pm 0.14$ &$-0.49\pm 0.13$ &$-0.57\pm 0.14$ &$-0.52\pm 0.15$ &$-0.58\pm 0.16$ &$-0.45\pm 0.12$ &$-0.4\pm 0.13$ &$-0.48\pm 0.13$ \\
Zn &$0.22\pm 0.12$ &$0.10\pm 0.11$ &$0.08\pm 0.14$ &$0.08\pm 0.12$ &$0.08\pm 0.15$ &$0.02\pm 0.11$ &$0.01\pm 0.14$ &$0.01\pm 0.12$ &$0.14\pm 0.15$ &$0.09\pm 0.16$ &$0.11\pm 0.12$ &$0.13\pm 0.12$ &$0.14\pm 0.11$ \\
Sr &$0.13\pm 0.13$ &$0.23\pm 0.11$ &$0.36\pm 0.12$ &$0.34\pm 0.14$ &$0.31\pm 0.11$ &$0.36\pm 0.13$ &$0.21\pm 0.14$ &$0.21\pm 0.15$ &$0.08\pm 0.13$ &$0.34\pm 0.12$ &$0.29\pm 0.13$ &$0.29\pm 0.11$ &$0.24\pm 0.12$ \\
Y &$-0.22\pm 0.13$ &$-0.32\pm 0.11$ &$-0.14\pm 0.10$ &$-0.24\pm 0.14$ &$-0.15\pm 0.10$ &$-0.1\pm 0.12$ &$-0.35\pm 0.14$ &$-0.17\pm 0.13$ &$-0.27\pm 0.12$ &$-0.02\pm 0.14$ &$-0.3\pm 0.11$ &$-0.13\pm 0.12$ &$-0.15\pm 0.14$ \\
Zr &$0.13\pm 0.12$ &$0.24\pm 0.14$ &$0.14\pm 0.12$ &$0.16\pm 0.15$ &$0.27\pm 0.11$ &$0.31\pm 0.13$ &$-0.06\pm 0.14$ &$0.16\pm 0.15$ &$0.18\pm 0.13$ &$0.19\pm 0.14$ &$0.11\pm 0.12$ &$0.24\pm 0.13$ &$0.25\pm 0.11$ \\
Ba &$-0.24\pm 0.09$ &$-0.06\pm 0.13$ &$0.08\pm 0.12$ &$-0.04\pm 0.14$ &$-0.02\pm 0.11$ &$0.15\pm 0.13$ &$-0.05\pm 0.10$ &$0.06\pm 0.11$ &$-0.15\pm 0.12$ &$0.04\pm 0.09$ &$-0.14\pm 0.11$ &$0.19\pm 0.12$ &$0.07\pm 0.10$ \\
La &$0.24\pm 0.10$ &$0.3\pm 0.12$ &$0.28\pm 0.11$ &$0.22\pm 0.11$ &$0.17\pm 0.09$ &$0.32\pm 0.10$ &$0.11\pm 0.13$ &$0.25\pm 0.12$ &$0.33\pm 0.14$ &$0.39\pm 0.10$ &$0.3\pm 0.11$ &$0.35\pm 0.08$ &$0.25\pm 0.13$ \\
Ce &$0.11\pm 0.11$ &$0.15\pm 0.11$ &$0.19\pm 0.13$ &$0.17\pm 0.10$ &$0.21\pm 0.09$ &$0.28\pm 0.13$ &$0.13\pm 0.10$ &$0.06\pm 0.12$ &$0.17\pm 0.15$ &$0.13\pm 0.10$ &$0.1\pm 0.12$ &$0.27\pm 0.11$ &$0.24\pm 0.14$ \\
Pr &$0.45\pm 0.12$ &$0.42\pm 0.15$ &$0.45\pm 0.14$ &$0.41\pm 0.11$ &$0.48\pm 0.12$ &$0.36\pm 0.14$ &$0.28\pm 0.13$ &$0.12\pm 0.12$ &$0.49\pm 0.14$ &$0.47\pm 0.15$ &$0.42\pm 0.08$ &$0.43\pm 0.13$ &$0.3\pm 0.11$ \\
Nd &$0.15\pm 0.15$ &$0.12\pm 0.12$ &$0.20\pm 0.13$ &$0.16\pm 0.14$ &$0.16\pm 0.11$ &$0.1\pm 0.13$ &$-0.1\pm 0.10$ &$0.02\pm 0.12$ &$0.19\pm 0.13$ &$0.29\pm 0.11$ &$0.17\pm 0.14$ &$0.19\pm 0.14$ &$0.14\pm 0.16$ \\
Sm &$0.36\pm 0.11$ &$0.25\pm 0.13$ &$0.32\pm 0.14$ &$0.34\pm 0.17$ &$0.25\pm 0.13$ &$0.21\pm 0.11$ &$0.11\pm 0.15$ &$0.13\pm 0.16$ &$0.37\pm 0.10$ &$0.11\pm 0.16$ &$0.22\pm 0.13$ &$0.27\pm 0.12$ &$0.31\pm 0.12$ \\
Eu &$0\pm 0.13$ &$-0.03\pm 0.12$ &$0.07\pm 0.14$ &$0.05\pm 0.15$ &$-0.03\pm 0.12$ &$0.04\pm 0.14$ &$-0.17\pm 0.15$ &$-0.18\pm 0.14$ &$0.09\pm 0.13$ &$0.07\pm 0.16$ &$-0.04\pm 0.14$ &$0.12\pm 0.12$ &$-0.05\pm 0.14$ \\
Gd &$0.45\pm 0.16$ &$0.38\pm 0.18$ &$0.50\pm 0.17$ &$0.21\pm 0.14$ &$0.49\pm 0.18$ &$0.46\pm 0.16$ &$999\pm 0.17$ &$0.14\pm 0.15$ &$0.33\pm 0.12$ &$-1.43\pm 0.19$ &$0.17\pm 0.16$ &$0.22\pm 0.21$ &$0.37\pm 0.17$\\
Dy &$0.24\pm 0.19$ &$0.16\pm 0.12$ &$0.53\pm 0.13$ &$0.23\pm 0.16$ &$0.48\pm 0.16$ &$999\pm 0.15$ &$-0.12\pm 0.18$ &$0.04\pm 0.22$ &$0.06\pm 0.16$ &$0.51\pm 0.13$ &$0.17\pm 0.15$ &$0.29\pm 0.16$ &$0.29\pm 0.18$ \\
Ho &$-0.52\pm 0.14$ &$0.18\pm 0.21$ &$0.62\pm 0.15$ &$-0.18\pm 0.15$ &$-0.05\pm 0.18$ &$-0.13\pm 0.14$ &$-0.29\pm 0.19$ &$-0.26\pm 0.16$ &$-0.13\pm 0.18$ &$999\pm 0.15$ &$0.36\pm 0.13$ &$-0.12\pm 0.19$ &$-0.28\pm 0.18$\\
Er &$0.28\pm 0.13$ &$0.32\pm 0.16$ &$0.33\pm 0.15$ &$0.17\pm 0.19$ &$0.51\pm 0.18$ &$0.29\pm 0.15$ &$-0.17\pm 0.23$ &$-0.21\pm 0.17$ &$0.21\pm 0.19$ &$0.14\pm 0.16$ &$0.04\pm 0.19$ &$0.19\pm 0.20$ &$0.22\pm 0.18$\\
Os &$0.13\pm 0.15$ &$0.26\pm 0.17$ &$0.42\pm 0.14$ &$0.26\pm 0.11$ &$-0.01\pm 0.15$ &$0.13\pm 0.14$ &$-0.15\pm 0.16$ &$0.09\pm 0.18$ &$-0.48\pm 0.12$ &$0.44\pm 0.14$ &$0.37\pm 0.15$ &$0.22\pm 0.17$ &$0\pm 0.16$\\
Th &$<-2.04$ &$<-1.26$ &$<0.29$ &$<-1.18$ &$<-0.33$ &$<-1.91$ &$<-1.22$ &$<-0.87$ &$<-1$ &$<-0.48$ &$<0.12$ &$<-1.01$ &$<-1.06$ \\
\enddata
\end{deluxetable}
\end{flushleft}
\label{fullabu}
\end{longrotatetable}

\startlongtable
\begin{deluxetable*}{l r l r r r c r c}
\tablewidth{100pt}
\tabletypesize{\footnotesize}
\tablecaption{\label{tab:linelist} Atomic Line Properties, Equivalent Widths, 
Absolute Abundances (before corrections), and Measurement Uncertainties of the Target Stars}
\tablehead{
\colhead{Star ID} & \colhead{Species} & \colhead{$\lambda$} & \colhead{$\chi$} & \colhead{$\log\,gf$} & \colhead{EW} & \colhead{$A(X)$}\\
& \colhead{({\AA})}      & \colhead{(eV)} &     &\colhead{(m{\AA})} &\colhead{(m{\AA})}   &  } 
\startdata
NGC2298-14 &NaI &5682.630 &2.100 &-0.710 &32.35 &4.735 \\
NGC2298-14 &NaI &5688.200 &2.100 &-0.410 &50.86 &4.739 \\
NGC2298-14 &NaI &5889.950 &0.000 &0.110 &297.23 &4.619 \\
NGC2298-14 &NaI &5895.920 &0.000 &-0.190 &259.50 &4.707\\ 
NGC2298-14 &NaI &6154.230 &2.100 &-1.550 &7.76 &4.826 \\
NGC2298-14 &NaI &8183.260 &2.100 &0.240 &120.49 &4.951 \\
NGC2298-14 &MgI &4167.270 &4.350 &-0.740 &76.52 &5.583 \\
NGC2298-14 &MgI &4702.990 &4.330 &-0.440 &106.40 &5.620 \\
NGC2298-14 &MgI &5528.400 &4.350 &-0.550 &118.02 &5.878 \\
NGC2298-14 &MgI &5711.090 &4.350 &-1.840 &31.95 &5.863 \\
NGC2298-14 &AlI &3944.000 &0.000 &-0.640 &315.98 &4.665 \\
NGC2298-14 &AlI &3961.520 &0.010 &-0.330 &409.01 &4.634 \\
NGC2298-14 &SiI &3905.520 &1.910 &-1.040 &279.67 &5.341 \\
NGC2298-14 &SiI &4102.940 &1.910 &-3.340 &105.99 &5.793 \\
NGC2298-14 &CaI &4318.650 &1.900 &-0.210 &95.96 &4.503 \\
NGC2298-14 &CaI &4425.440 &1.880 &-0.410 &86.21 &4.418 \\
NGC2298-14 &CaI &4455.890 &1.900 &-0.550 &82.02 &4.491 \\
NGC2298-14 &CaI &4456.610 &1.900 &-1.740 &25.86 &4.649 \\
NGC2298-14 &CaI &4578.550 &2.520 &-0.670 &41.53 &4.618 \\
NGC2298-14 &CaI &4685.270 &2.930 &-1.020 &20.54 &5.009 \\
NGC2298-14 &CaI &5261.710 &2.520 &-0.600 &51.76 &4.678 \\
NGC2298-14 &CaI &5512.980 &2.930 &-0.450 &28.0I4 &4.579 \\
NGC2298-14 &CaI &5581.970 &2.520 &-0.580 &51.89 &4.643 \\
NGC2298-14 &CaI &5588.760 &2.520 &0.300 &97.18 &4.566 \\
NGC2298-14 &CaI &5590.120 &2.520 &-0.600 &48.45 &4.608 \\
NGC2298-14 &CaI &5601.290 &2.530 &-0.570 &53.40 &4.669 \\
NGC2298-14 &CaI &5857.450 &2.930 &0.170 &75.82 &4.778 \\
NGC2298-14 &CaI &6102.720 &1.880 &-0.810 &96.64 &4.793 \\
NGC2298-14 &CaI &6122.220 &1.890 &-0.330 &128.18 &4.867\\
NGC2298-14 &CaI &6161.290 &2.520 &-1.350 &22.74 &4.840 \\
NGC2298-14 &CaI &6162.170 &1.900 &-0.110 &139.72 &4.834 \\
NGC2298-14 &CaI &6166.440 &2.520 &-1.220 &25.18 &4.766 \\
NGC2298-14 &CaI &6169.060 &2.520 &-0.870 &48.61 &4.847 \\
NGC2298-14 &CaI &6169.560 &2.530 &-0.600 &60.98 &4.788 \\
NGC2298-14 &CaI &6439.070 &2.520 &0.330 &116.22 &4.813 \\
NGC2298-14 &CaI &6471.660 &2.530 &-0.710 &51.53 &4.736 \\
NGC2298-14 &CaI &6499.650 &2.520 &-0.810 &48.46 &4.774 \\
NGC2298-14 &CaI &6717.690 &2.710 &-0.580 &56.82 &4.895 \\
NGC2298-14 &ScII &4246.820 &0.320 &0.240 &162.17 &1.412 \\
NGC2298-14 &ScII &5031.010 &1.360 &-0.400 &80.86 &1.256 \\
NGC2298-14 &ScII &5526.790 &1.770 &0.020 &76.92 &1.218 \\
NGC2298-14 &ScII &5657.910 &1.510 &-0.600 &70.33 &1.385 \\
NGC2298-14 &TiI &4512.730 &0.840 &-0.400 &56.58 &3.073 \\
NGC2298-14 &TiI &4533.240 &0.850 &0.540 &97.09 &2.946 \\
NGC2298-14 &TiI &4534.780 &0.840 &0.350 &81.92 &2.800 \\
NGC2298-14 &TiI &4548.760 &0.830 &-0.280 &55.67 &2.921 \\
NGC2298-14 &TiI &4555.480 &0.850 &-0.400 &54.29 &3.044 \\
NGC2298-14 &TiI &4656.470 &0.000 &-1.280 &73.00 &3.177 \\
NGC2298-14 &TiI &4681.910 &0.050 &-1.010 &80.80 &3.115 \\
NGC2298-14 &TiI &4840.870 &0.900 &-0.430 &51.35 &3.055 \\
NGC2298-14 &TiI &4981.730 &0.840 &0.570 &106.48 &3.002 \\
NGC2298-14 &TiI &4999.500 &0.830 &0.320 &102.86 &3.159 \\
NGC2298-14 &TiI &5016.160 &0.850 &-0.480 &57.03 &3.115 \\
NGC2298-14 &TiI &5173.740 &0.000 &-1.060 &89.02 &3.172 \\
NGC2298-14 &TiI &5192.970 &0.020 &-0.950 &106.87 &3.447 \\
NGC2298-14 &TiI &5210.380 &0.050 &-0.820 &102.17 &3.255 \\
NGC2298-14 &TiII &4028.340 &1.890 &-0.920 &88.02 &3.193 \\
NGC2298-14 &TiII &4394.060 &1.220 &-1.770 &90.07 &3.155 \\
NGC2298-14 &TiII &4395.840 &1.240 &-1.930 &84.86 &3.225 \\
NGC2298-14 &TiII &4399.770 &1.240 &-1.200 &118.43 &3.306 \\
NGC2298-14 &TiII &4417.710 &1.170 &-1.190 &121.18 &3.269 \\
NGC2298-14 &TiII &4418.330 &1.240 &-1.990 &86.87 &3.323 \\
NGC2298-14 &TiII &4443.800 &1.080 &-0.710 &142.93 &3.197 \\
NGC2298-14 &TiII &4444.550 &1.120 &-2.200 &89.00 &3.424 \\
NGC2298-14 &TiII &4470.850 &1.170 &-2.020 &88.81 &3.295 \\
NGC2298-14 &TiII &4493.520 &1.080 &-2.780 &52.94 &3.251 \\
NGC2298-14 &TiII &4501.270 &1.120 &-0.770 &150.06 &3.443 \\
NGC2298-14 &TiII &4583.410 &1.160 &-2.840 &43.35 &3.235 \\
NGC2298-14 &TiII &4636.320 &1.160 &-3.020 &31.06 &3.186 \\
NGC2298-14 &TiII &4708.660 &1.240 &-2.350 &68.79 &3.271 \\
NGC2298-14 &TiII &5129.160 &1.890 &-1.340 &85.84 &3.331 \\
NGC2298-14 &TiII &5185.900 &1.890 &-1.410 &77.00 &3.220 \\
NGC2298-14 &TiII &5336.790 &1.580 &-1.600 &92.27 &3.300 \\
NGC2298-14 &TiII &5381.020 &1.570 &-1.970 &77.56 &3.371 \\
NGC2298-14 &TiII &5418.770 &1.580 &-2.130 &65.81 &3.338 \\
NGC2298-14 &VI &4111.780 &0.300 &0.400 &71.40 &1.942 \\
NGC2298-14 &VI &4379.230 &0.300 &0.580 &81.20 &1.903 \\
NGC2298-14 &VI &4389.980 &0.280 &0.220 &66.06 &1.937 \\
NGC2298-14 &VII &4023.380 &1.800 &-0.610 &66.81 &2.180 \\
NGC2298-14 &CrI &4274.800 &0.000 &-0.220 &171.19 &3.505 \\
NGC2298-14 &CrI &4545.950 &0.940 &-1.370 &61.31 &3.450 \\
NGC2298-14 &CrI &4646.150 &1.030 &-0.740 &89.07 &3.468 \\
NGC2298-14 &CrI &4651.280 &0.980 &-1.460 &53.78 &3.443 \\
NGC2298-14 &CrI &5296.690 &0.980 &-1.360 &70.51 &3.554 \\
NGC2298-14 &CrI &5300.740 &0.980 &-2.000 &32.15 &3.547 \\
NGC2298-14 &CrI &5345.800 &1.000 &-0.950 &95.90 &3.646 \\
NGC2298-14 &CrI &5348.310 &1.000 &-1.210 &72.50 &3.458 \\
NGC2298-14 &CrI &5409.770 &1.030 &-0.670 &108.28 &3.648 \\
NGC2298-14 &CrII &4558.640 &4.070 &-0.430 &64.17 &3.829 \\
NGC2298-14 &CrII &4588.200 &4.070 &-0.650 &39.73 &3.574 \\
NGC2298-14 &CrII &4618.810 &4.070 &-0.890 &49.29 &3.998 \\
NGC2298-14 &MnI &4033.060 &0.000 &-0.650 &174.84 &3.411 \\
NGC2298-14 &MnI &4034.480 &0.000 &-0.840 &145.12 &3.096\\
NGC2298-14 &MnI &4041.350 &2.110 &0.280 &93.25 &3.308 \\
NGC2298-14 &MnI &4754.040 &2.280 &-0.080 &69.86 &3.184 \\
NGC2298-14 &MnI &4823.520 &2.320 &0.140 &91.45 &3.447 \\
NGC2298-14 &MnI &6021.820 &3.080 &-0.050 &TiII0 &3.155 \\
NGC2298-14 &MnI &3497.530 &1.850 &-1.420 &18.05 &1.420 \\
NGC2298-14 &FeI &4058.220 &3.210 &-1.180 &61.53 &5.536 \\
NGC2298-14 &FeI &4067.980 &3.210 &-0.530 &71.63 &5.090 \\
NGC2298-14 &FeI &4070.770 &3.240 &-0.870 &61.62 &5.260 \\
NGC2298-14 &FeI &4114.440 &2.830 &-1.300 &70.22 &5.367 \\
NGC2298-14 &FeI &4147.670 &1.490 &-2.070 &96.97 &5.104 \\
NGC2298-14 &FeI &4150.250 &3.430 &-1.190 &42.75 &5.448 \\
NGC2298-14 &FeI &4158.790 &3.430 &-0.700 &51.49 &5.118 \\
NGC2298-14 &FeI &4174.910 &0.910 &-2.940 &113.42 &5.667 \\
NGC2298-14 &FeI &4175.640 &2.850 &-0.830 &81.83 &5.174 \\
NGC2298-14 &FeI &4181.760 &2.830 &-0.370 &112.70 &5.457 \\
NGC2298-14 &FeI &4182.380 &3.020 &-1.180 &64.94 &5.359 \\
NGC2298-14 &FeI &4184.890 &2.830 &-0.870 &83.10 &5.216 \\
NGC2298-14 &FeI &4187.040 &2.450 &-0.560 &113.39 &5.111 \\
NGC2298-14 &FeI &4195.330 &3.330 &-0.490 &86.53 &5.528 \\
NGC2298-14 &FeI &4199.100 &3.050 &0.160 &114.05 &5.242 \\
NGC2298-14 &FeI &4217.550 &3.430 &-0.480 &81.19 &5.486 \\
NGC2298-14 &FeI &4222.210 &2.450 &-0.910 &100.97 &5.170 \\
NGC2298-14 &FeI &4233.600 &2.480 &-0.600 &114.92 &5.202 \\
NGC2298-14 &FeI &4238.810 &3.400 &-0.230 &78.37 &5.134 \\
NGC2298-14 &FeI &4250.120 &2.470 &-0.380 &132.89 &5.309 \\
NGC2298-14 &FeI &4282.400 &2.180 &-0.780 &114.45 &5.104 \\
NGC2298-14 &FeI &4388.410 &3.600 &-0.680 &54.73 &5.333 \\
NGC2298-14 &FeI &4430.610 &2.220 &-1.730 &95.65 &5.501 \\
NGC2298-14 &FeI &4442.340 &2.200 &-1.230 &118.95 &5.457 \\
NGC2298-14 &FeI &4447.720 &2.220 &-1.360 &110.97 &5.452 \\
NGC2298-14 &FeI &4466.550 &2.830 &-0.600 &124.12 &5.880 \\
NGC2298-14 &FeI &4484.220 &3.600 &-0.640 &53.61 &5.264 \\
NGC2298-14 &FeI &4494.560 &2.200 &-1.140 &126.77 &5.494 \\
NGC2298-14 &FeI &4547.850 &3.550 &-0.820 &51.33 &5.339 \\
NGC2298-14 &FeI &4592.650 &1.560 &-2.460 &123.06 &6.013 \\
NGC2298-14 &FeI &4595.360 &3.300 &-1.760 &37.47 &5.723 \\
NGC2298-14 &FeI &4602.000 &1.610 &-3.130 &64.82 &5.496 \\
NGC2298-14 &FeI &4602.940 &1.490 &-2.210 &115.73 &5.562 \\
NGC2298-14 &FeI &4607.650 &3.270 &-1.330 &51.58 &5.508 \\
NGC2298-14 &FeI &4619.290 &3.600 &-1.060 &41.75 &5.461 \\
NGC2298-14 &FeI &4630.120 &2.280 &-2.580 &55.90 &5.612 \\
NGC2298-14 &FeI &4637.500 &3.280 &-1.290 &53.27 &5.504 \\
NGC2298-14 &FeI &4643.460 &3.650 &-1.150 &39.76 &5.573 \\
NGC2298-14 &FeI &4647.430 &2.950 &-1.350 &76.78 &5.609 \\
NGC2298-14 &FeI &4661.970 &2.990 &-2.500 &17.91 &5.647 \\
NGC2298-14 &FeI &4668.130 &3.270 &-1.080 &65.37 &5.502 \\
NGC2298-14 &FeI &4704.950 &3.690 &-1.320 &23.67 &5.457 \\
NGC2298-14 &FeI &4733.590 &1.490 &-2.990 &92.55 &5.753 \\
NGC2298-14 &FeI &4736.770 &3.210 &-0.670 &88.79 &5.467 \\
NGC2298-14 &FeI &4800.650 &4.140 &-1.030 &16.73 &5.512 \\
NGC2298-14 &FeI &4807.710 &3.370 &-2.150 &19.11 &5.778 \\
NGC2298-14 &FeI &4871.320 &2.870 &-0.340 &118.61 &5.300 \\
NGC2298-14 &FeI &4872.140 &2.880 &-0.570 &121.88 &5.603 \\
NGC2298-14 &FeI &4875.880 &3.330 &-1.900 &23.60 &5.590 \\
NGC2298-14 &FeI &4882.140 &3.420 &-1.480 &34.37 &5.508 \\
NGC2298-14 &FeI &4885.430 &3.880 &-0.970 &32.49 &5.516 \\
NGC2298-14 &FeI &4890.760 &2.880 &-0.380 &132.56 &5.600 \\
NGC2298-14 &FeI &4891.490 &2.850 &-0.110 &132.63 &5.296 \\
NGC2298-14 &FeI &4903.310 &2.880 &-0.890 &106.11 &5.603\\
NGC2298-14 &FeI &4907.730 &3.430 &-1.700 &25.61 &5.556 \\
NGC2298-14 &FeI &4918.990 &2.860 &-0.340 &129.80 &5.482 \\
NGC2298-14 &FeI &4924.770 &2.280 &-2.110 &78.24 &5.512 \\
NGC2298-14 &FeI &4938.810 &2.880 &-1.080 &93.63 &5.529 \\
NGC2298-14 &FeI &4939.690 &0.860 &-3.250 &112.45 &5.588 \\
NGC2298-14 &FeI &4946.390 &3.370 &-1.110 &62.40 &5.561 \\
NGC2298-14 &FeI &4950.110 &3.420 &-1.500 &35.09 &5.536 \\
NGC2298-14 &FeI &4966.090 &3.330 &-0.790 &81.70 &5.551 \\
NGC2298-14 &FeI &4994.130 &0.920 &-2.970 &117.21 &5.478 \\
NGC2298-14 &FeI &5001.860 &3.880 &-0.010 &77.34 &5.357 \\
NGC2298-14 &FeI &5002.790 &3.400 &-1.460 &35.17 &5.470 \\
NGC2298-14 &FeI &5005.710 &3.880 &-0.120 &75.30 &5.438 \\
NGC2298-14 &FeI &5006.120 &2.830 &-0.620 &114.73 &5.425 \\
NGC2298-14 &FeI &5014.940 &3.940 &-0.180 &58.36 &5.251 \\
NGC2298-14 &FeI &5022.240 &3.980 &-0.330 &49.19 &5.291 \\
NGC2298-14 &FeI &5039.250 &3.370 &-1.520 &36.88 &5.523 \\
NGC2298-14 &FeI &5044.210 &2.850 &-2.020 &41.68 &5.473 \\
NGC2298-14 &FeI &5048.440 &3.960 &-1.000 &21.06 &5.376 \\
NGC2298-14 &FeI &5049.820 &2.280 &-1.360 &1CaI7 &5.609 \\
NGC2298-14 &FeI &5051.630 &0.920 &-2.760 &145.62 &5.848 \\
NGC2298-14 &FeI &5068.770 &2.940 &-1.040 &91.15 &5.490 \\
NGC2298-14 &FeI &5083.340 &0.960 &-2.840 &120.48 &5.445 \\
NGC2298-14 &FeI &5123.720 &1.010 &-3.060 &127.81 &5.877 \\
NGC2298-14 &FeI &5127.360 &0.920 &-3.250 &112.55 &5.617 \\
NGC2298-14 &FeI &5131.470 &2.220 &-2.520 &71.40 &5.691 \\
NGC2298-14 &FeI &5133.690 &4.180 &0.360 &76.40 &5.316 \\
NGC2298-14 &FeI &5141.740 &2.420 &-2.240 &64.74 &5.544 \\
NGC2298-14 &FeI &5150.840 &0.990 &-3.040 &116.38 &5.579 \\
NGC2298-14 &FeI &5166.280 &0.000 &-4.120 &131.63 &5.674 \\
NGC2298-14 &FeI &5171.600 &1.490 &-1.720 &136.22 &5.324 \\
NGC2298-14 &FeI &5191.450 &3.040 &-0.550 &128.93 &5.832 \\
NGC2298-14 &FeI &5192.340 &3.000 &-0.420 &122.21 &5.535 \\
NGC2298-14 &FeI &5194.940 &1.560 &-2.020 &130.55 &5.589 \\
NGC2298-14 &FeI &5198.710 &2.220 &-2.090 &90.69 &5.622 \\
NGC2298-14 &FeI &5202.340 &2.180 &-1.870 &112.20 &5.792 \\
NGC2298-14 &FeI &5215.180 &3.270 &-0.860 &77.04 &5.424 \\
NGC2298-14 &FeI &5216.270 &1.610 &-2.080 &126.79 &5.628 \\
NGC2298-14 &FeI &5217.390 &3.210 &-1.070 &67.44 &5.388 \\
NGC2298-14 &FeI &5225.530 &0.110 &-4.760 &99.49 &5.779 \\
NGC2298-14 &FeI &5232.940 &2.940 &-0.060 &138.01 &5.365 \\
NGC2298-14 &FeI &5242.490 &3.630 &-0.830 &46.30 &5.304 \\
NGC2298-14 &FeI &5253.460 &3.280 &-1.580 &36.74 &5.455 \\
NGC2298-14 &FeI &5263.310 &3.270 &-0.870 &78.61 &5.458 \\
NGC2298-14 &FeI &5266.560 &3.000 &-0.380 &119.43 &5.429 \\
NGC2298-14 &FeI &5281.790 &3.040 &-0.830 &92.11 &5.390 \\
NGC2298-14 &FeI &5283.620 &3.240 &-0.450 &107.42 &5.563 \\
NGC2298-14 &FeI &5288.520 &3.690 &-1.490 &15.49 &5.367 \\
NGC2298-14 &FeI &5307.360 &1.610 &-2.910 &82.96 &5.516 \\
NGC2298-14 &FeI &5322.040 &2.280 &-2.800 &41.84 &5.534 \\
NGC2298-14 &FeI &5324.180 &3.210 &-0.110 &119.98 &5.420 \\
NGC2298-14 &FeI &5332.900 &1.560 &-2.780 &89.84 &5.451 \\
NGC2298-14 &FeI &5339.930 &3.270 &-0.630 &92.44 &5.475 &\\
NGC2298-14 &FeI &5341.020 &1.610 &-1.950 &142.74 &5.783 \\
NGC2298-14 &FeI &5364.870 &4.450 &0.230 &65.77 &5.560 \\
NGC2298-14 &FeI &5365.400 &3.570 &-1.020 &43.67 &5.365 \\
NGC2298-14 &FeI &5367.470 &4.420 &0.440 &66.83 &5.332 \\
NGC2298-14 &FeI &5369.960 &4.370 &0.540 &76.04 &5.338 \\
NGC2298-14 &FeI &5373.710 &4.470 &-0.710 &15.93 &5.531 \\
NGC2298-14 &FeI &5379.570 &3.690 &-1.420 &17.61 &5.359 \\
NGC2298-14 &FeI &5383.370 &4.310 &0.640 &85.67 &5.346 \\
NGC2298-14 &FeI &5410.910 &4.470 &0.400 &56.94 &5.259 \\
NGC2298-14 &FeI &5415.200 &4.390 &0.640 &75.54 &5.249 \\
NGC2298-14 &FeI &5434.520 &1.010 &-2.130 &161.67 &5.497 \\
NGC2298-14 &FeI &5473.900 &4.150 &-0.720 &30.17 &5.510 \\
NGC2298-14 &FeI &5497.520 &1.010 &-2.820 &141.05 &5.792 \\
NGC2298-14 &FeI &5501.470 &0.960 &-3.050 &126.85 &5.675 \\
NGC2298-14 &FeI &5506.780 &0.990 &-2.790 &138.39 &5.682 \\
NGC2298-14 &FeI &5525.540 &4.230 &-1.080 &11.49 &5.444 \\
NGC2298-14 &FeI &5554.900 &4.550 &-0.270 &25.32 &5.432 \\
NGC2298-14 &FeI &5563.600 &4.190 &-0.750 &34.66 &5.671 \\
NGC2298-14 &FeI &5569.620 &3.420 &-0.520 &92.70 &5.525 \\
NGC2298-14 &FeI &5586.760 &3.370 &-0.110 &111.20 &5.405 \\
NGC2298-14 &FeI &5618.630 &4.210 &-1.250 &11.84 &5.600 \\
NGC2298-14 &FeI &5624.540 &3.420 &-0.760 &82.79 &5.573 \\
NGC2298-14 &FeI &5638.260 &4.220 &-0.720 &25.49 &5.486 \\
NGC2298-14 &FeI &5662.520 &4.180 &-0.410 &39.65 &5.410 \\
NGC2298-14 &FeI &5701.540 &2.560 &-2.140 &65.91 &5.585 \\
NGC2298-14 &FeI &5753.120 &4.260 &-0.620 &25.85 &5.437 \\
NGC2298-14 &FeI &5775.080 &4.220 &-1.080 &19.24 &5.684 \\
NGC2298-14 &FeI &5816.370 &4.550 &-0.600 &18.43 &5.575 \\
NGC2298-14 &FeII &4178.860 &2.580 &-2.510 &67.51 &5.051 \\
NGC2298-14 &FeII &4233.160 &2.580 &-2.020 &116.83 &5.648 \\
NGC2298-14 &FeII &4416.820 &2.780 &-2.570 &70.93 &5.386 \\
NGC2298-14 &FeII &4491.410 &2.860 &-2.710 &69.34 &5.581 \\
NGC2298-14 &FeII &4508.280 &2.860 &-2.420 &78.16 &5.467 \\
NGC2298-14 &FeII &4515.340 &2.840 &-2.600 &71.15 &5.480 \\
NGC2298-14 &FeII &4576.340 &2.840 &-2.950 &54.98 &5.509 \\
NGC2298-14 &FeII &4583.830 &2.810 &-1.940 &113.44 &5.667 \\
NGC2298-14 &FeII &5197.580 &3.230 &-2.220 &65.70 &5.382 \\
NGC2298-14 &FeII &5234.630 &3.220 &-2.180 &74.66 &5.518 \\
NGC2298-14 &FeII &5276.000 &3.200 &-2.010 &90.52 &5.581 \\
NGC2298-14 &FeII &5325.550 &3.220 &-3.160 &21.58 &5.473 \\
NGC2298-14 &CoI &3894.080 &1.050 &0.120 &173.28 &3.825 \\
NGC2298-14 &CoI &3995.310 &0.920 &-0.180 &109.30 &2.753 \\
NGC2298-14 &CoI &4121.320 &0.920 &-0.330 &109.13 &2.849 \\
NGC2298-14 &NiI &4604.990 &3.480 &-0.240 &35.16 &4.098 \\
NGC2298-14 &NiI &4648.650 &3.420 &-0.090 &44.77 &4.048 \\
NGC2298-14 &NiI &4686.220 &3.600 &-0.590 &11.14 &3.948 \\
NGC2298-14 &NiI &4714.420 &3.380 &0.250 &71.92 &4.149 \\
NGC2298-14 &NiI &4756.520 &3.480 &-0.270 &34.75 &4.108 \\
NGC2298-14 &NiI &4904.410 &3.540 &-0.170 &44.04 &4.240 \\
NGC2298-14 &NiI &4980.170 &3.600 &0.070 &48.64 &4.148 \\
NGC2298-14 &NiI &5017.580 &3.540 &-0.030 &41.37 &4.045 \\
NGC2298-14 &NiI &5080.530 &3.650 &0.320 &58.49 &4.121 \\
NGC2298-14 &NiI &5081.110 &3.850 &0.300 &38.73 &4.036 \\
NGC2298-14 &NiI &5084.080 &3.680 &0.030 &37.59 &4.081 \\
NGC2298-14 &NiI &5115.400 &3.830 &-0.110 &28.62 &4.224 \\
NGC2298-14 &NiI &6643.630 &1.680 &-2.220 &80.24 &4.452 \\
NGC2298-14 &NiI &6767.770 &1.830 &-2.140 &69.66 &4.382 \\
NGC2298-14 &CuI &5105.500 &1.390 &-1.500 &35.30 &1.681 \\
NGC2298-14 &CuI &4722.160 &4.030 &-0.370 &37.76 &2.619\\
NGC2298-14 &ZnI &4810.540 &4.080 &-0.150 &47.14 &2.630 
\enddata
\tablecomments{The linelist for one star is shown here. The full table for all the targets will be made available electronically.}
\end{deluxetable*}
\facilities{Gemini South: GHOST}
\end{document}